\documentclass[fleqn,usenatbib]{mnras}
\usepackage{newtxtext,newtxmath}
\DeclareRobustCommand{\VAN}[3]{#2}
\let\VANthebibliography\thebibliography
\def\thebibliography{\DeclareRobustCommand{\VAN}[3]{##3}\VANthebibliography}

\newcommand{\G}{\textit{Gaia}}

\newcommand{\nobj}{598}%
\newcommand{\nsys}{278}%
\newcommand{\double}{241}
\newcommand{\triple}{33}
\newcommand{\higherorder}{4}
\newcommand\gbp{\ensuremath{G_\mathrm{BP}}}
\newcommand\grp{\ensuremath{G_\mathrm{RP}}}
\newcommand\grvs{\ensuremath{G_\mathrm{RVS}}}
\newcommand\g{\ensuremath{G}}
\newcommand\gaia{\textit{Gaia}}
\newcommand{\nucd}{4530}
\newcommand{\nsysadapted}{41}
\newcommand{\nadapted}{46}

\newcommand{\asymerr}[3]{%
  {#1}^{\raisebox{0.3ex}{\scriptsize${{#2}}$}}_{\raisebox{-0.3ex}{\scriptsize${{#3}}$}}%
}

\usepackage{booktabs}
\usepackage{graphicx}	
\usepackage{amsmath}	
\usepackage{siunitx}    
\usepackage{tabularx}   
\usepackage{makecell}   
\usepackage{float}
\usepackage{caption}
\usepackage[utf8]{inputenc}
\usepackage{gensymb} 
\usepackage{lscape} 
\usepackage{adjustbox} 
\usepackage{multirow} 
\usepackage{aas_macros}
\usepackage{hyperref} 
\usepackage{subcaption}
\usepackage{tablefootnote}
\usepackage{enumitem} 
\usepackage{calc}
\usepackage{longtable}
\usepackage{xtab}
\usepackage{ltablex}

\title[GUCDS V: UCD Companions]{The  \G~Ultracool
  Dwarf Sample -- V: The Ultracool Dwarf Companion catalogue} 

\author[Sayan Baig]{
Sayan Baig$^{1,2}$,\thanks{E-mail: s.baig@herts.ac.uk}
R. L. Smart$^{1,2}$,
Hugh R.A. Jones$^{1}$,
Jonathan Gagné$^{3,4}$,
D.J. Pinfield$^{1}$,
Gemma Cheng$^{1}$,
\newauthor \phantom{} Leslie Moranta$^{3,4,5}$
\\
$^{1}$ School of Physics, Astronomy and Mathematics, University of Hertfordshire, College Lane, Hatfield, AL10 9AB, UK\\
$^{2}$ Istituto Nazionale di Astrofisica, Osservatorio Astrofisico di Torino, Strada Osservatorio 20, I-10025 Pino Torinese, Italy\\
$^{3}$ Planétarium Rio Tinto Alcan, Espace pour la Vie, 4801 av. Pierre-de Coubertin, Montréal, Québec, Canada\\
$^{4}$ 
Institute for Research on Exoplanets, Universit\'e de Montr\'eal, D\'epartement de Physique, C.P. 6128 Succ. Centre-ville, Montr\'eal, QC H3C 3J7, Canada\\
$^{5}$ Department of Astrophysics, American Museum of Natural History, Central Park West at 79th St., New York, NY 10024, USA
}

\date{Accepted XXX. Received YYY; in original form ZZZ}

\pubyear{2024}

\begin{document}
\label{firstpage}
\pagerange{\pageref{firstpage}--\pageref{lastpage}}
\maketitle

\begin{abstract}

We present the Ultracool Dwarf Companion Catalogue of 278 multiple systems, 32 of which are newly discovered, each with at least one spectroscopically confirmed Ultracool Dwarf, within a 100 pc volume-limited sample. This catalogue is compiled using the $\gaia$ Catalogue of Nearby Stars for stellar primaries and the $\gaia$ Ultracool Dwarf Sample for low-mass companions and includes $\double$
doubles, $\triple$ triples, and $\higherorder$ higher-order systems established from positional, proper motion, and parallax constraints. 
The catalogue seeks to identify probable benchmark systems within 100\,pc to obtain model-independent astrophysical parameters of Ultracool Dwarfs.
Chance alignment probabilities are calculated to evaluate the physical nature of each system. Astrometric and photometric data from $\gaia$ Data Release 3 and the Two Micron All Sky Survey are included for all objects. We identify potential unseen companions using a combination of the Renormalised Unit Weight Error, Image Parameter Determination statistics, Non-Single Star solutions, and photometric blending as provided by $\gaia$, identifying hierarchical Ultracool triple systems. Our catalogue includes 17 White Dwarf - Ultracool Dwarf systems, whose ages are determined using cooling models. We also use the Gaia FLAME results and the BANYAN $\Sigma$ procedures to age 40 and 34 systems respectively, and derive mass estimates from evolutionary models.

\end{abstract}

\begin{keywords}
binaries: general – brown dwarfs – stars: low-mass
\end{keywords}



\section{Introduction}
$\gaia$ Data Release 3 
\citep[DR3,][]{2023A&A...674A...1G}
was made publicly available on June 13, 2022, building on the Gaia Early Data Release 3 \citep[EDR3,][]{eDR3} which contains astrometric solutions (parallax, sky position ($\alpha$, $\delta$), and proper motions) for 1.468 billion sources, with a limiting magnitude of $\g$ $\sim$ 21\,mag and a brightness limit of $\g$ $\sim$ 3\,mag. Advances in EDR3 have helped in the search for real binary systems as the median uncertainties in parallaxes and one-dimensional proper motions have improved from 0.165\,mas to 0.120\,mas and from 0.280\,mas yr$^{-1}$ to 0.123\,mas yr$^{-1}$ respectively, at $\g$ = 18\,mag \citep{EDR3_astrometry}. The improvements in accuracy are more significant at brighter magnitudes, with the median parallax uncertainty at $\g$ = 13\,mag improving from 0.029 to 0.015\,mas and proper motion precision improved by a factor of 2 \citep{2021MNRAS.506.2269E}. This improved precision enables the distinction of widely separated binary systems at distances greater than what is possible by using Gaia Data Release 2 \citep[DR2, ][]{GaiaDR2}

DR3 enhances the capabilities of EDR3 by including object classifications for 1.8 billion sources, with approximately 470 million sources having derived astrophysical parameters from low-resolution BP/RP spectra \citep{2023A&A...674A..27A}, such as the effective temperature (T$_{\rm{eff}}$), surface gravity ($\log\,g$), metallicity ([M/H]), age, and distance, determined from low-resolution $\grp$/$\gbp$ ($\g$ $<$ 18.25 mag) and spectra from the Radial Velocity Spectrometer (RVS) ($\g$ $<$ 15 mag). One of the key features of DR3 is the inclusion of Non-Single Star (NSS) solutions with $\sim$ 813,000 sources, of which approximately 170,000 have astrometrically derived acceleration solutions (non-linear proper motion). This provides a unique opportunity to independently estimate the masses of unseen companion stars in binary systems, particularly for cooler and fainter companions, that have previously been challenging to observe directly.

Ultracool Dwarfs (UCDs) are cool (T$_{\rm{eff}}$ $\le$ 2700K) low-mass objects that straddle the stellar substellar mass boundary (0.075 $M_{\odot}$). UCDs are defined as objects with an M7 or later spectral type, which includes both low-mass stars and Brown Dwarfs (BDs) \citep{kirkpatrick_1997, Rajpurohit_2013}, and thus defines the boundary between stellar hydrogen-burning stars and degenerate BDs \citep{ 2019MNRAS.485.4423S}.  
Observable UCDs are located at close distances because of their intrinsic faintness, with a limited number expected to be detected by the $\gaia$ mission, with most L-dwarfs observed no further than 80 pc away \citep{2019MNRAS.485.4423S}. Consequently, our study is confined to within 100 pc to ensure the inclusion of the majority of potentially detectable UCDs, aligning with the observational capabilities of $\gaia$. Although $\gaia$ may not be an ideal instrument for studying low-mass, faint, red stars, its precision in trigonometric parallax and proper motion, combined with the vast amount of data available from DR3, provide an excellent opportunity to study and establish a catalogue of UCD companion systems. 
The study of low-mass UCDs has gained prominence in recent years,  and an early example is the discovery of the T-dwarf Gl229B, which was found to be in a companion system with an early M-dwarf \citep{1995_Gl229B}. This discovery led to the determination of the system properties, as reported by \cite{Gl229B_properties}. 

Multiple-star systems provide valuable information for stellar physics because the fundamental physical properties of one component can be used to infer the properties of other components \citep{Serenelli_2021}. Astrophysical properties are difficult to measure for low-mass objects due to their faintness; however, UCDs in multiple systems with well-defined Main Sequence (MS) primary companions (brighter in the $\g$-band) can constrain the properties of secondary UCD objects, such as their composition and age \citep{2006MNRAS.368.1281P}. By assuming coevality, the degeneracy between the mass and age of BDs can also be broken \citep{2010AJ....139..176F}. Recent advances in astronomical technologies have led to a notable increase in the number of directly imaged UCDs that have masses determined independently from evolutionary models. These benchmark systems aid in the evaluation and refinement of prevailing stellar and substellar models \citep{2013ApJ...771...46C, 2015ApJ...798L..43C, 2016ApJ...831..136C, 2020ApJ...904L..25C, 2021ApJS..254...42B,2022AJ....163..288C, 2023MNRAS.522.5622L}, spectral synthesis, and atmospheric retrievals \citep{2022AJ....163..189W}.\\

Our understanding of BDs has been largely informed by theoretical modelling progress, with iterative refinements of existing models and the introduction of novel substellar models to capture the intricacies of BD evolution \citep{1995ApJ...446L..35B, 1996Sci...272.1919M, 2002A&A...382..563B, 2008ApJ...689.1327S, 2011ApJ...736...47B}. Post-formation, BDs cool over several million years following a mass-luminosity-age relationship, which forms the backbone of the aforementioned evolutionary models.

A key issue in employing substellar models arises from the large modelling uncertainties. These uncertainties are particularly pronounced in measurements of age, luminosity, and [Fe/H], posing significant barriers to the effective use of evolutionary models and BD cooling theories. Examination of BD benchmarks with substellar models highlights underestimated luminosities of young BDs as seen with HD 130948BC \citep{2009ApJ...692..729D} and overestimated luminosities for older BDs such as HD 4113C shown by \cite{2018A&A...614A..16C}. The spectra of UCDs display strong alkaline absorption lines and broad molecular absorption bands that are influenced by surface gravity, metallicity and effective temperature \citep{2009AJ....137.3345C}. Despite attempts to establish a correlation between these spectral features and UCD properties, this relationship suffers from a significant level of dispersion and a limited sample size. These inconsistencies underpin the necessity of an expanded sample of UCD benchmark systems to facilitate a more exhaustive evaluation of the existing cooling models.

The utilisation of wide-field surveys, such as the Two Micron All Sky Survey (2MASS) \cite{Two_Mass}, Sloan Digital Sky Survey (SDSS) \citep{1999RSPTA.357...93M}, the Wide-field Infrared Survey Explore (WISE) \citep{2010AJ....140.1868W}, The UKIRT Infrared Deep Sky Survey (UKIDSS) \citep{2007MNRAS.379.1599L}, The Visible and Infrared Survey Telescope for Astronomy (VISTA) \citep{2015A&A...575A..25S} and the first Panoramic Survey Telescope and Rapid Response System release \citep[PS1, ][]{2012ApJ...750...99T} has led to the discovery of new UCDs and the characterisation of their binarity, as demonstrated in \cite{2000ApJ...531L..57B, 2009MNRAS.395.1237B, Burningham_tdwarf, 2010AJ....139..176F, 2010MNRAS.404.1817Z, 2011MNRAS.410..705D, 2012ApJ...760..152L(luwe19),     2012MNRAS.422.1922P, 2013MNRAS.431.2745G, 2020MNRAS.499.5302D}. The advent of $\gaia$ has greatly increased the scope for identifying benchmark UCDs binary systems; \cite{2017MNRAS.470.4885M} estimates 2,960 resolvable UCD benchmarks from $\gaia$ alone, identifying 13 new benchmark UCD systems in their analysis. \newline
This study aims to assist in establishing reliable ages and masses of these elusive objects while increasing the current sample size of benchmark UCDs. The Ultracool Dwarf Companion Catalogue (hereafter UCDC) opens new avenues for exploring the formation variances between MLTY dwarfs compared with more massive stars, constraining the Present-day Mass Function, spatial distribution, and binary fraction in the wider context of galactic star formation and evolution.

The remainder of this paper is organised as follows: In Sec.\ref{cat creation} we discuss the creation of the UCDC, calculation of the False Positive Probability and draw comparisons with The Fifth Catalogue of Nearby Stars (CNS5). Sec.\ref{indentify binaries} describes the use of the Renormalised Unit Weight Error (RUWE), blended photometry, and $\gaia$ NSS solutions to identify close-compact UCD binary systems. Sec.\ref{mass_age_section} describes the estimation of age and mass from  White Dwarf (WD) - UCD binaries, the $\gaia$ Final Luminosity Age Mass Estimator (FLAME) and BANYAN $\Sigma$ \citep{2019yCat..18560023G}. Finally, we present a summary and final remarks in Sec.\ref{conclusion}.

\section{Catalogue creation}
\label{cat creation}
\subsection{The Gaia Ultracool Dwarf Sample}

The Gaia Ultracool Dwarf Sample (GUCDS) developed in \cite{2019MNRAS.485.4423S} provides a catalogue of over 20,000 objects spanning spectral types M7-M9, L, T, and Y, as well as companion objects. The spectral types for approximately 80$\%$ of these objects were spectroscopically confirmed, estimates for the remainder were derived from photometry. The GUCDS collates photometric information from various surveys by incorporating data from 2MASS, PS1, WISE, and $\gaia$, if available. In addition, GUCDS supplements UCD entries with astrometry, mainly from $\gaia$ or ground-based surveys. Roughly 25$\%$ of UCDs within the GUCDS are fainter than the $\gaia$ detection limit and thus the inclusion of ground-based observational data of UCDs beyond $\gaia$'s detection limit provides a larger pool of UCDs companions. To assemble the UCD sample for this study, the GUCDS was restricted to spectroscopically confirmed UCDs within 100 pc ($\varpi \ge$ 10 mas, within 3$\sigma$ error), yielding $\nucd$ selected UCDs.

\subsubsection{UCD spectral classifications}
An initial catalogue that included photometric and spectrally classified UCDs was produced using the GUCDS (see Sec.\ref{inital search} for details on the catalogue creation process). We performed a TAP query to cross-match the initial catalogue with the SIMBAD TAP service \citep{1991ASSL..171...79E} using a 10$\arcsec$ search limit to account for discrepancies in position. SIMBAD typically provides spectral types along with a bibcode reference, if available for an object, along with a quality letter (ranging from A to E, with A being the best). We cross-matched our sample of UCDs to identify any with photometric spectral classifications in the GUCDS that now have spectral classifications available for inclusion in the final catalogue. We found that the spectrally classified UCDs in the GUCDS were generally in good agreement with SIMBAD entries, with variations in spectral types primarily due to incorrect object matching and slight discrepancies in subclasses resulting from differences in the selected studies. However, these differences were not significant for the UCD status of catalogue members. 

Seven UCDs with photometric spectral types in the GUCDS are identified with a spectral type from this search. SIMBAD does not clarify whether the spectral typing listed is spectrally or photometrically based; thus, a literature review of these 7 objects was conducted, revealing 4 UCDs having true spectral classifications, with the remaining objects being photometrically estimated.

Four additional UCDs were spectrally confirmed, resulting in four new companion UCDs: 
\begin{itemize}
\item \textit{{SDSS J154005.12+010208.8}}: Classified as an M9 in \cite{2019AJ....157..231K} with optical spectra listed as 'brighterL' from the SDSS skyserver \cite{2001cs.......11015S}. 
\item \textit{{SDSS J155738.27+335602.1}}: Classified as M9e \citep{2019AJ....157..231K} with the SDSS optical spectra as 'fainterL'. 
\item \textit{{SDSS J144633.50+363126.1}}: Classified as an M9 \citep{2019AJ....157..231K}, and listed as an L-dwarf from the SDSS optical spectra. 
\item \textit{{2MASS J15104761-2818234}}: classified as M9 in \citep{2002ApJ...575..484G}, in further agreement with \cite{2020NatAs...4..650T} as M9 in both the optical and NIR spectra. 
\end{itemize}

Three UCDs were rejected because of the following photometric classifications:
\begin{itemize}
\item \textit{{2MASS J01015311+1528195}}: Listed as an M9.5 in \cite{2011EPJWC..1606014Z}, but it does not have an SDSS spectrum and its spectral type has been photometrically estimated from SDSS colours.
\item \textit{{SDSS J091817.14+264037.2}}: Classified as an M7 in \cite{2015AJ....149..158S} through photometric estimation of its spectral type based on SDSS colours.
\item \textit{{2MASS J13171150+1849232}}: Given a photometrically estimated spectral type of M9V in \cite{2011AJ....141...97W} without an available SDSS spectrum.
\end{itemize}

\subsubsection{GCNS}
The purpose of the $\gaia$ Catalogue of Nearby Stars \citep[GCNS, ][]{GCNS} is to provide a high-quality catalogue of nearby objects within a 100 pc radius of the Sun, utilising data from EDR3. The catalogue includes all $\gaia$ sources with $\varpi$ $>$ 8mas, with spurious objects systematically removed using a random forest classifier and Bayesian distance probability function. Distance-limited samples often contain contamination from distant sources with unreliable parallaxes due to poor astrometric fitting, for which the random forest classifier is effective at removing, as shown in Fig. 1 of \cite{GCNS}. The catalogue includes 339,312 sources, with completeness expected to be approximately 95$\%$ for objects up to spectral type M7 and decreasing completeness for later spectral types with L8 objects only complete up to 10\,pc. The GCNS magnitude distribution peaks at $\g$ $\sim$ 20.4\,mag and only contains objects that are identified in $\gaia$. Faint objects that are not detected by $\gaia$ are therefore missed in the GCNS but included in the GUCDS, which includes ground-based observations of faint stars and photometrically identified objects, such as those described by \cite{2016A&A...589A..49S}.

The GCNS is a useful resource for studying binary systems, as it provides a concise and reliable sub-sample of the entire $\gaia$ catalogue that operates at similar distances to the UCDs that are directly observable. 
Nearby stars in the GCNS typically have well-defined astrophysical parameters, which make them potentially ideal as primaries in benchmark systems.

\subsection{Initial Catalogue}
\label{inital search}
We initially searched for UCD companion systems between the GUCDS and GCNS by adopting the criteria used by \cite{2019MNRAS.485.4423S}. The criteria consist of four separate cuts, and all systems must satisfy the conditions to be considered as a companion.

\begin{itemize}
    \item \textbf{Projected Separation}: The angular separation (\arcsec) between the two candidates, $\rho$, must satisfy: \\
    \begin{equation}
    \rho(\arcsec) < 100\varpi(\text{mas})
     \end{equation}
where $\varpi$ is the parallax (mas) of the UCD. The formulation of $\rho$ corresponds to a physical separation of 100,000\,AU, which is a conservative upper limit for the possible projected separation (s). Wide binaries are subject to very small (absolute) gravitational potential energies. The separation limit will ensure the binding energy criterion is met, $U_g = GM_1M_2/s$ $>$ 10$^{33}$J, for a 0.1+2M$_{\odot}$ binary system \citep{2009A&A...507..251C,2010AJ....139.2566D}. It is expected that the occurrence of real binaries is significantly less likely beyond this distance, as the galactic tidal field becomes proportional to the gravitational attraction between the two candidates. The point at which the tidal field becomes stronger than the gravitational attraction is the Jacobi radius $r_j$ (full derivation can be found in \citealt{Jiang}).
A more liberal criterion was implemented by \cite{2021MNRAS.506.2269E}, taking the limit of $\rho$ to allow for physical separations between components to be as large as 1\,pc, which is roughly double our limit. However, the study was not limited to just systems with a UCD component. \\

\item \textbf{Parallax}:  The difference in parallaxes $\varpi_{1}$ and $\varpi_{2}$ (represents the primary and secondary respectively, both in mas) must satisfy: 
\begin{equation}
    \Delta\varpi < \max\left[1.0,3\sqrt{\sigma_{\varpi_1}^2 + \sigma_{\varpi_2}^2}\right]
\end{equation}

where $\Delta\varpi$ is the difference between the parallaxes and $\sigma_{\varpi_1}$ and $\sigma_{\varpi_2}$ are the errors of the primary and secondary objects, respectively. In general, we require a 3$\sigma$ consistency; however, if $\Delta\varpi$ $<$ 1, a maximum difference threshold of 1\,mas is used. For the GUCDS, the median parallax error across all UCDs is approximately 1 mas, justifying the selection of this value. \cite{2021MNRAS.506.2269E} shows a correlation between underestimated parallax errors and poor astrometric fitting from $\gaia$ by considering the Renormalised Unit Weight Error (RUWE) and Image Parameter Determination (IPD) quantities, most notably from sources with $\g$ $>$ 13 as they are fit with a 1D Line Spread Function (LSF) thus inherently leading to biases for poor astrometric fitting for close sources.    \\

\item \textbf{Proper Motion}: The proper motions of the two candidates in a wide binary are expected to have similar values; however, they are not identical because of the effects of orbital motions. To account for significant orbital motion, $\mu$ (proper motion of the UCD) should be within 10\% of the difference in the total proper motion.
\begin{equation}
    \Delta\mu < 0.1\mu
\end{equation}
where $\Delta\mu$ is the difference in the total proper motion, determined by 
\begin{equation}
    \Delta\mu = [(\mu_{\alpha,1} - \mu_{\alpha,2})^2 + (\mu_{\delta,1} - \mu_{\delta,2})^2) ]^{1/2}
\end{equation}
where $\delta$ and $\alpha$ denote the right ascension and declination, respectively, and the proper motions in the right ascension and declination directions are denoted by $\mu_{\alpha}$ and $\mu_{\delta}$. It should be noted that  $\mu_{\alpha}$ = $\mu_{\alpha}\cos\delta$; which is the local tangent plane projection of the proper motion vector in the direction of increasing right ascension, DR3 data includes this in the raw value, and thus does not need to be accounted for. \\

\item \textbf{Direction of Proper Motion}: We assume that binary systems have common proper motions and any slight differences in their direction would be due to orbital motion. To account for this, a 15$^{\circ}$ tolerance is applied, as follows: 
\begin{equation}
    \Delta\theta < 15^{\circ} 
\end{equation}
where $\Delta\theta$ is the difference between the proper motion direction. 
\end{itemize}

As outlined in Sec.2.3 of \cite{2019MNRAS.485.4423S}, there are acknowledged shortcomings in using these binary criteria. One plausible reason for these shortcomings is the significant contribution of the orbital motion of the system to the objects' proper motion, resulting in a proper motion difference that exceeds the initially set 10$\%$ tolerance. It is also appropriate to highlight that the criteria initially applied to DR2 data may not be directly translatable to DR3 data.

\subsection{Adaptation of binary criteria }
\label{adapted_cuts}
We reflect on the choices made in our initial search (Sec.\ref{inital search}) with an example of the well-established K3 + L1.5 companion system GJ 1048A + GJ 1048 B \citep{2001AJ....121.2185G} which does not meet our binary criteria because the proper motion difference exceeds the initial 10$\%$ tolerance limit at $\sim$ 13$\%$. 
The large proper motion (PM) discrepancy indicates that the system is sufficiently close and slow enough that the orbital motions are significant to the contribution of the PM.  We note that \cite{2021MNRAS.506.2269E} provides a different approach to the PM criteria, describing that wide binary systems should be consistent with Keplerian orbits, however, employing this method still results in the failure of this system to pass our binary criteria. 

\begin{figure}
\includegraphics[width=\columnwidth]{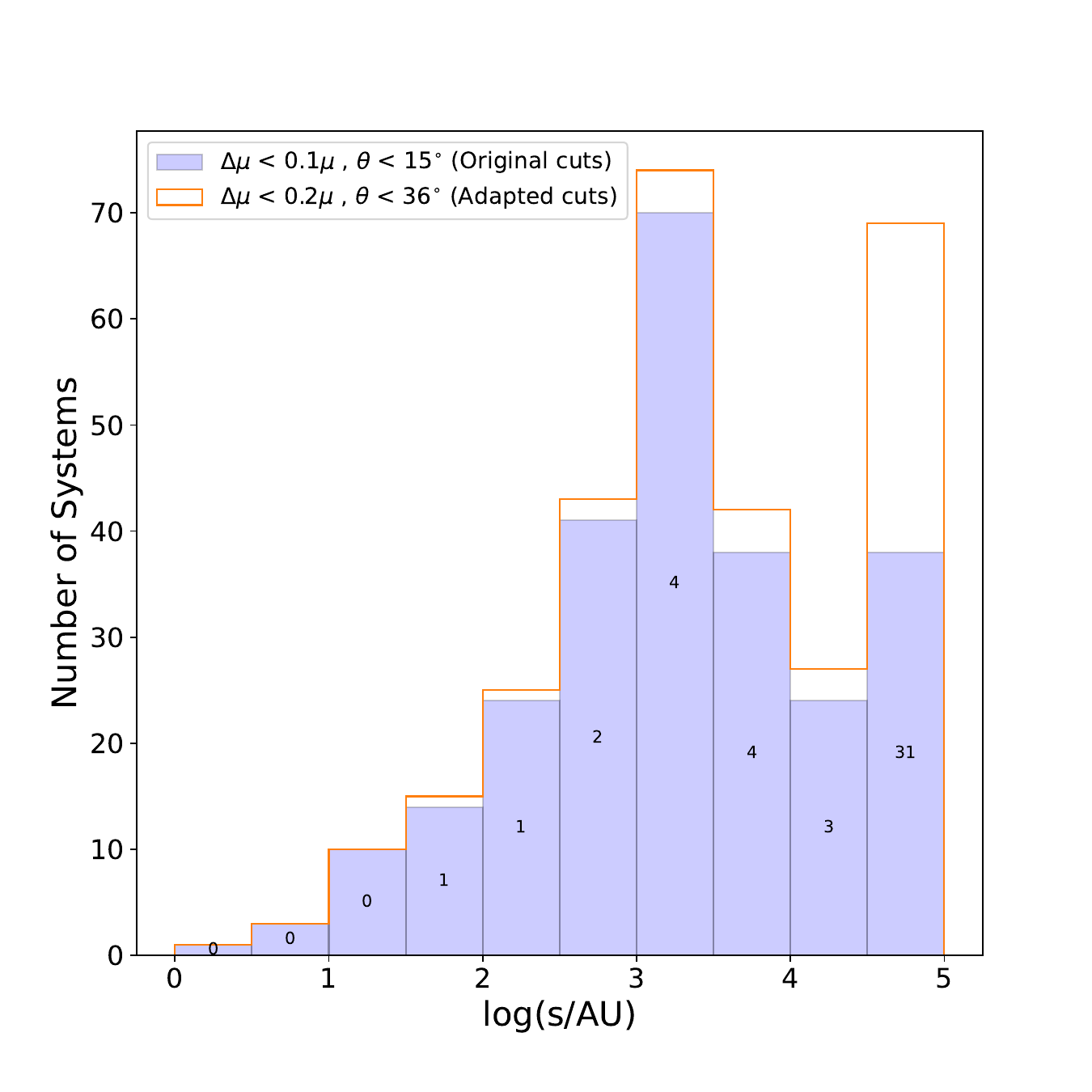}
\caption{Distribution of the number of systems by the log of projected physical separation (s) between constituents. The blue histogram represents the original cuts described in Sec.\ref{inital search} and the gold unfilled histogram represents the adapted cuts described in Sec.\ref{adapted_cuts}. The difference in systems per bin is represented in the centre of each histogram bar.}
\label{fig:hist_of_sep}
\end{figure}

This study aims to create a detailed catalogue of UCD companion systems. Thus, we relax the proper motion difference to 20\% ($\Delta\mu < 0.2\mu$) to include GJ 1048A + GJ 1048 B and other potentially missed systems. To maintain consistency with the PM amplitude, the PM direction tolerance is increased to $ \Delta\theta$ $<$ 36$^{\circ}$. \newline
The adapted criteria produce an additional $\nsysadapted$ systems from $\nadapted$ newly included objects. We have 3 additional systems that could have significant orbital motions as they have separations of $<$ 1000\,AU. The inclusion of all new systems is due to the adjusted PM tolerance, as none of these systems exceeded the original directional tolerance specified in the initial criteria. The total systems per separation bin for both criteria is displayed in Fig. \ref{fig:hist_of_sep}. The adapted criteria distribution closely follows the initial criteria except for log(s/AU) = 4.5 ($\sim$ 30,000\,AU). \cite{2021MNRAS.506.2269E} explains that contamination rates increase with separation, and binary separation distributions decrease over the same separation range. The increase in binary candidates at log(s$/$AU) = 4.5 marks the point at which these contaminants dominate the sample. In this case, an additional 31 systems are included with separations of log(s$/$AU) $\ge$ 4.5. Given the leniency of the criteria, additional spurious systems may be included, coupled with new real systems. It should be noted that 13 systems with log(s/AU) $\ge$ 4.5 are either triples or higher-order multiples. A distant tertiary companion can be included at these separations in regions with high stellar density. However, as \cite{2021MNRAS.506.2269E} elucidates, there is a high probability of chance alignment (optical doubles) at these distances, and it is predicted that removing resolved triples and moving groups would result in a maximum loss of 15$\%$ of real multiples, as there are indeed bound systems at these separations. It was decided that resolved triples, higher-order systems, and systems within moving groups would remain in the catalogue to prevent the loss of genuine multiples. 
\\

Despite the more liberal criteria, known UCD systems still fail to make the UCDC, owing to significant orbital motion effects and large parallax differences. We are aware of two systems which are listed in Table \ref{failures_newcuts}. HD 212168 and CPD-75 1748 B form a companion system \citep{2010ApJS..190....1R} which was not included because the PM difference between the two systems is $\approx$ 43$\%$. We also find HD 212168 to be in a wide companion with an M8 UCD, DENIS J222644.3-750342, suggesting a missed triple system due to the omission of CPD-75 1748 B.\\
TYC-3424-215-1 and TYC-3424-215-2 form a wide binary as suggested by \cite{Tycho-Gaia} with an angular separation $\rho$ = 6.09\arcsec which fails to satisfy the parallax criterion. TYC-3424-215-1 and TYC-3424-215-2 are identified as wide companions to  2MASS J09073765+4509359 \citep{2019MNRAS.485.4423S, superwide, Kervella_EDR3_companions}. We find only TYC 3424-215-1 and 2MASS J09073765+4509359 as a double system ($\varpi$ = 26.99 $\pm$ 0.12\,mas and $\varpi$ = 26.74 $\pm$ 0.27\,mas); however, TYC 3424-215-2 has a significantly larger parallax ($\varpi$ = 31.77 $\pm$ 0.66\,mas) and thus is not included.

\subsubsection{False Positives}
False Positives (FPs) are inherent in statistical binary samples arising from chance alignments with an increasing False Positive Probability (FPP) as the separation between companions increases \citep{2021MNRAS.506.2269E}. Detailed consideration of the FPP must be performed to ensure a robust sample of physically associated companion systems. We compute the FPP for each system and present the methodology below. 

We first consider the probability that a star in the GCNS has an angular separation ($\rho$) comparable to that of the true companion ($\rho_\text{{comp}}$) in our sample (P$_{\text{prox}}$). We find the total number of objects with $\rho$ $\le$ $\rho_\text{{comp}}$ (N$_\text{{sep}}$) in the GCNS and the total number of objects in the GCNS within 100,000\,AU (N$_\text{{100k}}$) to the primary. To mitigate the bias of FPs arising from genuine physical companions, we have excluded all companion stars of the primary objects in each system within the GCNS. The GCNS provides a table of resolved binary companions which we cross-reference with the UCDC, identifying 150 sources. Subsequently, any additional matches are categorised as FPs.

\begin{equation}
P_{\text{prox}} = \frac{N_{\text{sep}}}{N_{\text{100k}}}
\end{equation}

There are instances in which no object in the GCNS has $\rho$ $\le$ $\rho_\text{{comp}}$ (i.e., $N_\text{{sep}}$ = 0). This is typical for the tightest binary separations or areas where GCNS objects are sparse. To resolve instances of $N_\text{{sep}}$ = 0 we uniformly expand the parameter space to allow a larger separation. This is achieved by proportionally scaling both $N_\text{{sep}}$ and $N_\text{{100k}}$, for example, considering 5 times the initial separation \textbf{(5$N_\text{{sep}}$/5$N_\text{{100k}}$)}. This scaling was required for 34 systems, accounting for $\sim$ 12\% of the catalogue.

This scaling approach is premised on the uniformity of surface density within the GCNS, as suggested by the distribution of selected sources in Fig 1 of \cite{2021A&A...649A...6G}. However, it is acknowledged that localised non-uniformities exist, such as the over-density in the Hyades and the Galactic plane due to their larger stellar density and structural complexity. Consequently, the scaling of parameters, although uniform, may lead to slightly conservative estimations of true companions in the plane and conversely, be marginally optimistic for less populated areas. \\

The probability a star within 100,000\,AU has a similar parallax of the primary is calculated as : 
\begin{equation}
P_{\text{plx}} = \frac{N_{\text{plx}}}{N_{\text{100k}}}
\end{equation}
$N_{\text{plx}}$ is the relative number of objects with a parallax within 3$\sigma$ of the primaries parallax. To address instances where the number of stars within the original 3$\sigma$ parallax range is insufficient ($N_{\text{plx}}$ = 0), we adapt our approach by expanding the area of the sky we consider. This method involves searching for common parallax objects at further distances, for instance, extending the search from 100,00\,AU (N$_\text{{100k}}$) to 200,000\,AU(N$_\text{{200k}}$). This increased range allows us to consider a broader set of stars that are still statistically significant in terms of their parallax similarity to the primary companion. This expansion is based on the assumption of a constant parallax distribution across the sky, which suggests that the relative distribution of parallax values does not significantly vary with the position of the stars, thus an effective way to capture a more representative sample without distorting the underlying distribution given the non-uniformity in parallax distribution.

A strong covariance exists between parallax and PM, combining $P_{\text{plx}}$ and the probability of having a star with a similar PM to the primary ($P_{\text{pm}}$) into a single probability calculation may not accurately reflect the individual influences of parallax and PM. To resolve this issue, we separate the calculation of $P_{\text{pm}}$ by considering only those objects within $N_{\text{plx}}$. By limiting our analysis to objects within $N_{\text{plx}}$, we increase the likelihood that similarities in PM reflect a genuine physical association, reducing the potential confounding effects of similar motions observed at varying distances. The PM similarity is then assessed within this subset:
\begin{equation}
P_{\text{pm}} = \frac{N_{\text{pm}}}{N_{\text{plx}}}
\end{equation}
\( N_{\text{pm}} \) represents the number of stars within the 3$\sigma$ parallax range that also have a total proper motion within 20\% of the primary target. It should be noted that as the PM criterion is based on the system's total PM, faster-moving objects will naturally exhibit a larger range in their 20\% PM allowance, consequently leading to increased detection of FPs.

We also evaluate the probability of the direction of the PM (P$_{\text{dir}}$) for each object within N$_\text{{100k}}$, where N$_{\text{dir}}$ represents the total objects with a PM direction $<$ 36$^{\circ}$ or  $>$ 324$^{\circ}$ compared with the primary target: \begin{equation}
P_{\text{dir}} = \frac{N_{\text{dir}}}{N_{\text{100k}}}
\end{equation} 
We note that cases of \(N_{\text{pm}} = 0\) and \(N_{\text{dir}} = 0\) are treated in a manner analogous to instances of \(N_{\text{plx}} = 0\).
\\

We consider the hypothesis (H) that a candidate binary is a FP, i.e. a line of sight association whose GCNS measurements (or evidence E) are consistent with binarity.
Thus the fraction of which FPs could arise in the GCNS parameter space is :
\begin{equation}
P_{\text{GCNSFP}} = P_{\text{prox}} \times P_{\text{plx}} \times P_{\text{pm}} \times P_{\text{dir}}
\end{equation}

To calculate the FPP, $P(H|E)$, we apply Bayes Theorem :
\begin{equation}
P(H|E) = \frac{P(H)P(E|H)}{P(H)P(E|H) + P(\neg H)P(E|\neg H)}    
\end{equation}
where:
\begin{itemize}
    \item \( P(E|H) \), denoted as \( P_{\text{GCNSFP}} \), is the likelihood function, representing the likelihood of a chance alignment given the evidence ($P_{\text{prox}}$, $P_{\text{plx}}$, $P_{\text{pm}}$, $P_{\text{dir}}$).
    \item \( P(\neg H) \) is given as \( P_{\text{comp}} \times \frac{1}{N_{100k}} \), signifying the prior probability that the primary has a companion, multiplied by the probability of selecting the correct companion from the initial sample of candidate companions. We choose to be P$_{\text{comp}}$ = 0.075 from the UCD binary fraction in \cite{2019ApJ...883..205B}. 
    \item \( P(H) \) is \( 1 - P(\neg H) \), the prior probability for a false positive before evidence is taken into account. 
    \item \( P(E|\neg H) \) is the probability of observing the specific evidence given that the hypothesis of chance alignment is false, and the detected companion is a true physical associate of the primary. This probability is set to 1.0, reflecting the certainty that such evidence will be present in every instance where the companion is a genuine physical companion. The evidence is not merely coincidental but is an expected signature of physical binary or multiple-star systems.
\end{itemize}

To calculate the probability of the UCDC containing a FP, we first determine the probability of having no FPs in the sample, \( P(\rm{FP} = 0) \).

Let \( p_i \) denote the probability of the \(i\)-th system being a FP. The probability of the \(i\)-th system not being a FP is \( 1 - p_i \). The probability of having no FPs in the entire catalogue (N) is then given by the product of the probabilities that each individual system is not a FP:
\[
P( \rm{FP} = 0) = \prod_{i=1}^{N} (1 - p_i)
\]

this product yields \( P(\rm{FP} = 0) = 0.33 \).

The probability of having a FP in the sample is computed by subtracting the probability of having no FPs from 1:
\[
 P(\rm{FP}) = 1 -  P(\rm{FP} = 0) = 1 - 0.33 = 0.67
\]

Additionally, we calculate the cumulative number of FPs, (P$_{\text{CFP}}$), which is the sum of the individual FP probabilities:
\[
P_{\text{CFP}} = \sum_{i=1}^{N} p_i
\]

\begin{figure}
\includegraphics[width=\columnwidth]{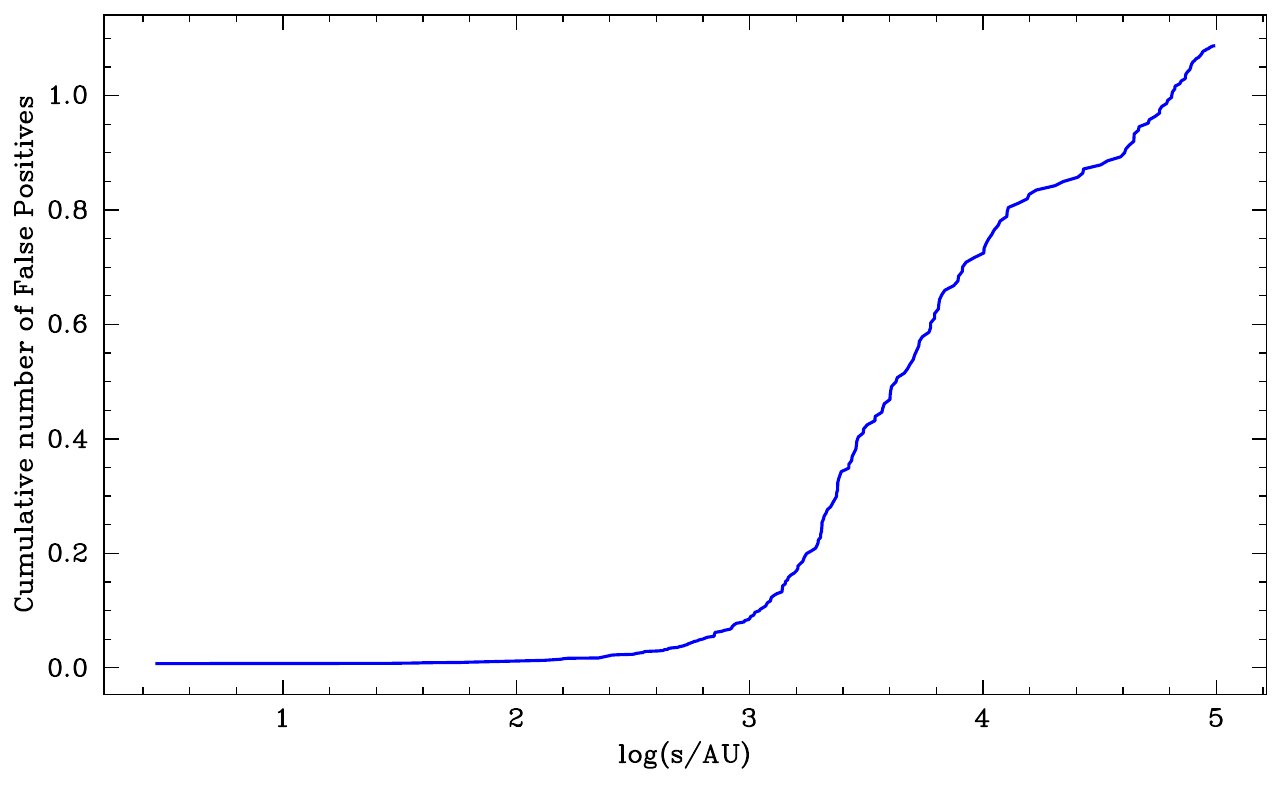}
\caption{Cumulative number of False Positives (P$_{\text{CFP}}$) as a function of binary separation.  }

\label{fig:cdf}
\end{figure}

In this case, P$_{\text{CFP}}$ is 1.09 as depicted in Fig. \ref{fig:cdf}, illustrating a progressive increase in the cumulative likelihood of encountering a FP as a function of the logarithmic projected separation, 
log(s). This observed trend is attributable to an increased presence of background objects as binary separation increases. The true binary separation distribution is known to decrease monotonically with increasing binary separation \citep{2018MNRAS.480.4884E}. Given the comparatively smaller binding energies characteristic of UCD companion systems, it is reasonable to anticipate that a FP would be more likely at smaller separations compared to what is reported by \cite{2021MNRAS.506.2269E} which examines all binary systems irrespective of mass, observing a steep increase in Cumulative Distribution Function (CDF) of FPs at log(s) = 4.5. This stems from the inherent physical properties of UCD systems, where smaller binding energies imply a higher susceptibility to disruption at lower separation, thereby altering the CDF-FP profile when compared to binary systems of all mass ranges. In Fig.\ref{fig:cdf}, log(s) $\sim$ 3 marks the point at which FPs become more probable and notably at log(s) = 4.5 there is a steep ascent in the likelihood of a FP. This inflexion point signifies where the contamination of the catalogue is likely to dominate. The alignment of our results with those presented by \cite{2021MNRAS.506.2269E} reinforces the interpretation that FPs are more prevalent at larger separations, and this should be a consideration in the analysis and vetting of candidates in such catalogues.

\subsubsection{The 200pc sample}

Initially, the UCDC was compiled within a 200\,pc distance limit, however, while late M-Dwarfs are observable with $\gaia$ over this distance, their detection is limited by the $\gaia$ magnitude limit as completeness for late M-dwarfs falls beyond 80\,pc \citep{GCNS}.
We opt to use the GCNS as we know it is complete up to 100\,pc and the distance limit is similar to the distance at which M-dwarfs are complete in $\gaia$.
The reduction in the distance limit has resulted in the loss of some genuine UCD companion systems, such as the known hierarchical quadruple system 2MASS J04414565+2301580 AabBab \citep{2015ApJ...811L..30B}. Other companions in nearby Star Forming Regions (SFRs) are invariably excluded from the UCDC as the focus is to construct a local sample out to 100\,pc and not to be complete out to local SFRs which exist beyond 100\,pc, as discussed in \cite{2018ApJ...856...23G}.

\subsection{Known UCD companion systems not identified in the UCDC}

One of the challenges in the study of UCD companion systems is the lack of astrometric data. $\gaia$, while providing comprehensive 5-parameter astrometric solutions is limited to objects with $\g$ $<$ 21\,mag, leaving many fainter UCDs without the astrometric solutions necessary to investigate potential binarity. Additionally, resolution limitations are also a factor to consider as $\gaia$ is unable to resolve equal-mass systems below $\sim$ 0.2 $\arcsec$ \citep{2023A&A...674A...1G}.  

To mitigate these limitations, the UCDC was established using primarily $\gaia$-derived astrometric data, supplemented by additional objects, provided in the GUCDS. Despite these measures, there exists a selection of UCDs companion systems, known from direct imaging and characterisation in the existing literature, that were initially overlooked in the UCDC due to the lack of independent or $\gaia$ based astrometric solutions. These systems are presented in Table \ref{no astrometry table} and have been included in the catalogue by adopting the astrometric solution of their primary companion. 

\subsection{Comparisons with the CNS5}
The Fifth Catalogue of Nearby Stars  \citep[CNS5, ][]{2023A&A...670A..19G} is a comprehensive inventory of all nearby objects up to a 25 pc volume limit, including basic astrometric and photometric parameters. The catalogue contains parallaxes from infrared ground-based surveys and 541 non-$\gaia$ UCDs from  \cite{2021AJ....161...42B} and \cite{2021ApJS..253....7K}. The CNS5 contains 5931 objects, complete up to $\g$ $\sim$ 19.7 mag, including a total of 701 UCDs. As the CNS5 distance limit overlaps with the UCDC and both are assembled using similar procedures, a comparison is useful to determine any discrepancies in UCD companion systems. A comparison of the UCDC and CNS5 was conducted using a 10$\arcsec$ radius to account for the positional variances of non-$\gaia$ objects due to differences in epochs. The cross-match yielded 127 matches, including 26 non-$\gaia$ objects whereas the UCDC contains 144 objects with $\varpi$ $\ge$  40 mas.  A notable characteristic of the CNS5 is its inclusion of only visual binaries, as spectroscopic systems have been excluded from the catalogue. This selective inclusion accounts for the discrepancy in the number of matched objects within 25\,pc.

\subsubsection{Omissions from the CNS5}
Unlike the UCDC, the CNS5 employs a binary criterion that omits a parallax constraint and adopts an alternative proper motion condition. 
 A comparison with the UCDC reveals 20 objects that are not classified as part of a companion system in the CNS5 but feature in the UCDC, some of which belong to systems found spectroscopically and thus not considered to be discrepant. The visual binaries identified in the UCDC, but not in the CNS5, are listed in Table \ref{cns5_table}. The individual stars are included in the CNS5; however, they are not recognised as part of a companion system.
The CNS5's proper motion constraint, as used by \cite{2021MNRAS.506.2269E}, excludes these multiples as bona fide Common Proper Motion (CPM) systems. 

Equation 3 in \cite{2023A&A...670A..19G} imposes a constraint on the CNS5 which requires objects to have a parallax consistent with 40\,mas within 3$\sigma_{\varpi}$ and $\sigma_{\varpi}$ $\le$ 10mas. This restriction results in the exclusion of possible UCD companions from the CNS5, as they are not considered to be within 25 pc with sufficient reliability. The UCDC includes  J0102+0355, an L9 UCD \citep{2016ApJ...817..112S} along with a wide companion, J0102+0355A, with a separation of $\rho$ = 9.87$\arcmin$. Although J0102+0355 has a $\varpi$ = 40.2 $\pm$ 2.8 mas, satisfying the 25 pc criterion, J0102+0355A has a $\varpi$ = 34.99 $\pm$ 0.037, which falls outside of this range and consequently has been excluded from the CNS5. The uncertainty in the parallax measurement for J0102+0355 allows the system to remain in the UCDC despite the large difference in the parallax.
Two triple systems listed in the UCDC are not included in the CNS5, J2201+3222 and J1927-4833. In both cases, the companions of these UCDs lie beyond 25 pc.
Finally, J1112+3548BC is unresolved in $\gaia$ consisting of a L4.5+L6 in a wide triple system with J1112+3548 \citep{2001AJ....121.3235K}. Although J1112+3548BC is known to the CNS5 and is mentioned in Table A.1 of \cite{2023A&A...670A..19G}, it is not included because of the constraints imposed by Equation 2 in \cite{2023A&A...670A..19G}, which aims to eliminate false astrometric solutions.
\\

\subsubsection{Omissions from the UCDC}
To locate the UCD binaries in the CNS5, a limit of M$_G$ $\ge$ 13.54\,mag was implemented, following the photometric relations of \cite{2019AJ....157..231K} for an M7. This was necessary because spectroscopic classifications were not available for all objects in the CNS5. Applying this criterion yielded a subset of 108 objects, of which 33 were present in the UCDC. The remaining objects were excluded due to a lack of available spectra or non-UCD classifications.

Three UCD companion systems are included in the CNS5, but not in the UCDC.  Two of the systems, Smethells 79, and its companion UCAC4 238-179131 and G 9-38AB fail due to large parallax differences. Although the systems have similar positions and space motions, the parallax difference between the constituents of each system exceeds the 3$\sigma$ tolerance in Equation 2 of Sec.\ref{inital search}. G 9-38AB, which consists of an M7v+M7v pair \citep{2014AJ....147...20N}, was first discovered in \cite{1985A&A...148..151P}.

2MASSI J2249091+320549 (J2249+3205) is listed in the CNS5 as part of a binary system with a non-$\gaia$ object, with the companion's astrometry sourced from \cite{2020AJ....159..257B}. No companions for J2249+3205 were identified in either the GUCDS or the GCNS, explaining its absence in the UCDC. The absence of any indications of binarity, such as a resolved double or partially resolved Point Spread Function (PSF) in high-resolution images from the Wide-Field Near-Infrared Camera (WFCAM) mounted on the United Kingdom Infrared Telescope (UKIRT) \citep{UKIDSS}, suggests that J2249+3205 may indeed be a single star. This hypothesis is further supported by the lack of mention of the binarity for J2249+3205 in \cite{2020AJ....159..257B}.  Furthermore, follow-up observations utilising the Near Infrared Camera and Multi-Object Spectrometer (NICMOS) on the Hubble Space Telescope \citep{2022AJ....164..244F} likewise provided no evidence of binarity for J2249+3205. We therefore posit J2249+3205 is not in a binary system given the evidence discussed.

\subsection{The Ultracool Dwarf Companion catalogue}
\label{UCDC section}
\begin{figure*}
\includegraphics[clip,width=\textwidth]{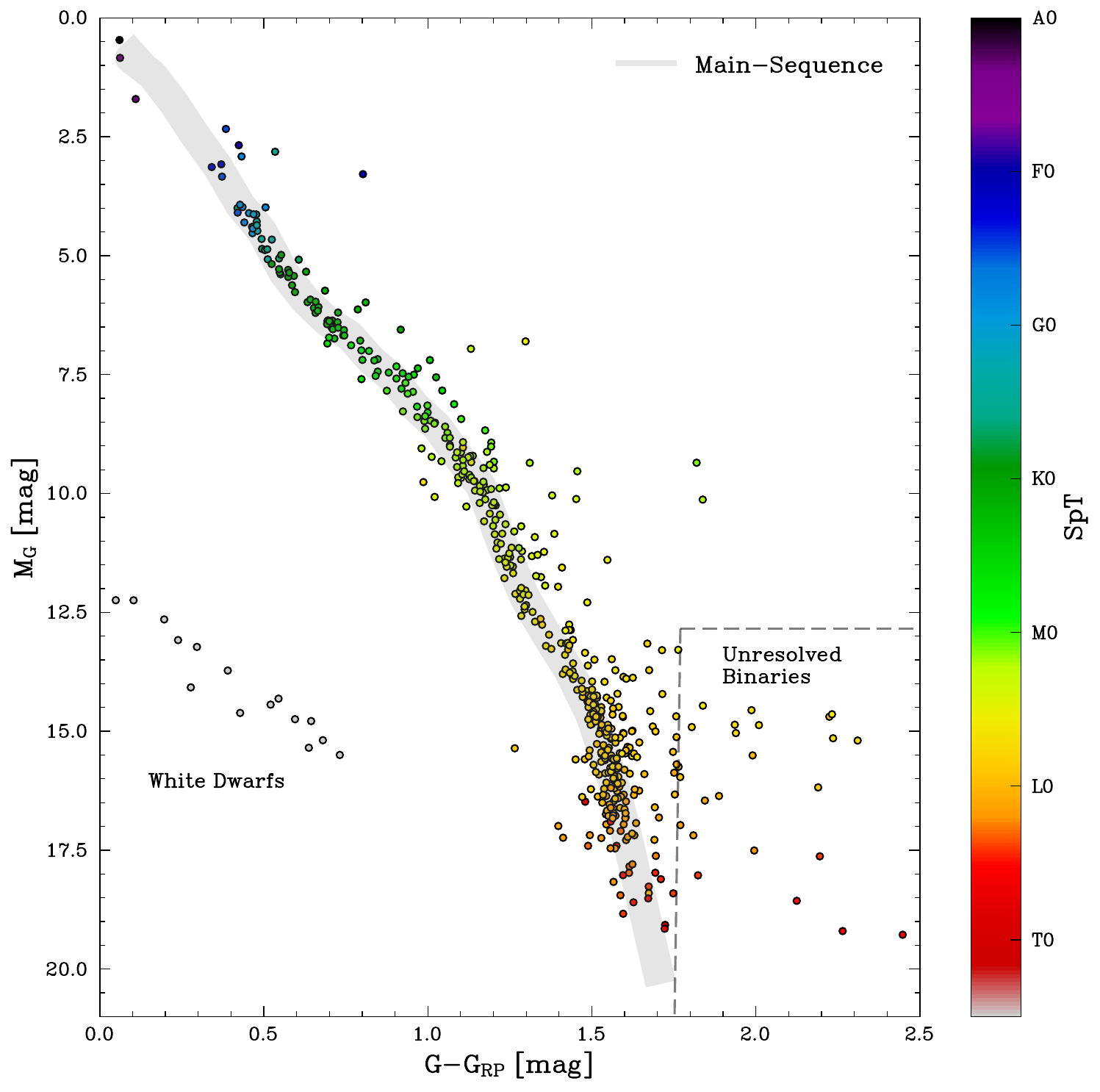}
\caption{CMD of the UCDC catalogue in G-$\grp$. Each point on the plot represents a single object, with the colour of the point indicating the spectral type. 
Objects that have G-$\grp$ $\ge$ 1.77 and M$_G$ $\ge$ 12.84 are defined as 'Unresolved Binaries' (explained in Sec.\ref{blended section}). Note: The 59 non-$\gaia$ objects in the UCDC are not included in this figure as no $\gaia$ photometry exists for them. The spectral types provided by the GUCDS have been used for this figure, of which some non-UCD objects are photometrically estimated.}
\label{fig:cmd}
\end{figure*}

The UCDC comprises $\nobj$ objects, including $\double$ double systems, $\triple$ triple systems, and $\higherorder$ systems with four or more components (higher-order), making $\nsys$ systems in total. Of these systems, 32 are newly discovered including 29 doubles and 3 triple systems.  An example of a system from the catalogue is presented in Table \ref{Table1}. For brevity, Table \ref{Table1} does not display the error columns;  the complete catalogue includes errors and can be accessed online. Fig \ref{fig:cmd} presents the Colour Magnitude Diagram (CMD) of the UCDC along with the Main Sequence (MS), which is primarily an approximate representation used in this and subsequent figures. A running median of the absolute magnitude was computed across discrete colour bins. To present this running median with a functional form, a polynomial fitting technique was used, with a 6th-degree polynomial tailored to the median data: 

\begin{equation}
M_G = \sum_{i=1}^{6} a_i (G - G_{\text{RP}})^i + 1.903
\end{equation}
where a$_1$  = 22.47,  a$_2$  = -149.2,  a$_3$  = 300.5,
a$_4$  = -291.5, a$_5$  = 135.9, \\ a$_6$ \,=\,24.7.

Although it offers an illustrative insight, it is worth noting that the provided polynomial equation is merely an estimation.\\

While this catalogue represents a concerted effort to capture a significant subset of UCD multiple systems, it is important to consider the caveats of the UCDC. The catalogue's dependence on a manual literature review to search for UCD multiple systems lacking astrometric constraints means that there are potential real systems missed or close companions resolved in the literature not included. We identify real systems that fail our binary criteria from a literature search, indicating the possibility of more systems also missed due to the limitations of the literature search. Our UCD sample is solely from the GUCDS, meaning that any UCDs within 100\,pc that are not found in either the GCNS or GUCDS are consequently missed.

\begin{table*}
\centering
\caption{Known systems that have failed the adapted cuts described in \ref{adapted_cuts} DN - Discovery Name, ID - $\gaia$ source ID}
\begin{tabular}{llllllll }
\toprule DN & ID  &  $\g$ & $\varpi$ & $\mu_\alpha$ & $\mu_\delta$ & $\mu$ 
 \\  &     &  (mag) & (mas) & (mas yr$^{-1}$) & (mas yr$^{-1}$) & (mas yr$^{-1}$) \\ \midrule 
HD 212168 & 6357835694518769408 & 5.98 & 42.72  & 57.39 & 12.83 & 57.39
\\
CPD-75 1748B & 6357835488360338560 & 8.38 & 42.69 & 33.33 & -3.79 & 33.33
\\
\hline
TYC 3424-215-2 & 1009609617150473600 & 9.48& 31.77 & 74.48 & -29.63 & 80.15
\\
2MASS J09073765+4509359 & 1009608650782545280 & 18.98 & 26.74 & 68.57 & -36.74 & 77.79
\\ \bottomrule
 \end{tabular}
\label{failures_newcuts}
\end{table*}

\begin{table*}
\caption{Companion systems with constituents lacking astrometric data that have been included in the UCDC. Binarity Reference  - bibliographic code associated with the initial discovery of the binary nature. SN - Short Name }
\label{no astrometry table}
\begin{tabular}{@{}lllllll@{}}
\toprule
SN & DN &  ID & $\varpi$ & $\mu_\alpha$ & $\mu_\delta$  \\ 
& & & (mas) &   (mas yr$^{-1}$) & (mas yr$^{-1}$) \\ \midrule
J0027+2239A$^1$& LP 349-25 & 2799992744809482112 & 70.78 $\pm$ 0.43 & 392.72 $\pm$ 0.47 & -186.59 $\pm$ 0.40   \\
J0027+2239B & LP 349-25B & - & - & - & - & \\
\hline
J0122-2439A$^2$ & 2MASS J01225093-2439505 & 5040416186560252416 & 29.64 $\pm$ 0.03 & 120.22 $\pm$ 0.03 & -123.56 $\pm$ 0.02  \\
J0122-2439 & 2MASS J01225093-2439505b  & - & - & - & - &  \\
\hline
J1106+2754$^3$ & 2MASS J11061197+2754225 & 731881226310429056 & 49.27 $\pm$ 1.22 & -270.81 $\pm$ 1.24 & -451.85 $\pm$ 1.22   \\
J1106+2754B & 2MASS J11061197+2754225B  & - & - & - & - &  \\
\hline
J1121-1313B$^4$ & LP 732-94 & 3562157781229213312 & 69.49 $\pm$ 0.18 & -472.25 $\pm$ 0.33 & -46.26 $\pm$ 0.21  \\
J1121-1313 & LHS2397aB  & - & - & - & - &  \\
\hline
J1217+1427A$^5$ & HIP 59933 & 3921176983720146560 & 15.45 $\pm$ 0.08 & -103.15 $\pm$ 0.09 & -37.55 $\pm$ 0.07 \\
J1217+1427 & 2MASS J12173646+1427119 & 3921177219942653696 & - & - & - &  \\
\hline
J1256-1257A$^6$ & 2MASS J12560215-1257217 & 3526198184723289472 & 47.27 $\pm$ 0.47 & -272.46 $\pm$ 0.57 & -190.24 $\pm$ 0.50   \\
J1256-1257 & VHS J1256-1257B  & - & - & - & - &  \\
\hline
J1324-5130A$^7$ & HIP 65426 & 6070080754075553792 & 9.30 $\pm$ 0.03 & -33.92 $\pm$ 0.03 & -18.92 $\pm$ 0.03  \\
J1324-5130 & HIP 65426 b  & - & - & - & - &  \\
\hline
J1423+0116A$^8$ & HD 126053 & 3654496279558010624 & 57.27 $\pm$ 0.04 & 223.53 $\pm$ 0.05 & -478.28 $\pm$ 0.03\\
J1423+0116 & ULAS J142320.79+011638.2 & - & - & 224.00 $\pm$ 0.57 & -477.75 $\pm$ 0.42 &  \\
\hline
J1454+1606$^9$ & BD+16 2708 & 1187851653287128576 & 100.7 $\pm$ 0.05 & 279.34 $\pm$ 0.08 & -117.96 $\pm$ 0.10  \\
J1454+1606B & BD+16 2708b & 1187851653287128576 & - & - & - &  \\
\hline
J1735+2634$^{10}$ & LP 388-55 & 4594186745414679680 & 64.33 $\pm$ 0.75 & 141.58 $\pm$ 0.63 & -289.13 $\pm$ 0.83  \\
J1735+2634B & LP 388-55B & - & - & - & - &  \\
\hline
J2004+1704A$^{11}$ & 15 Sge & 1821708351374312064 & 56.27 $\pm$ 0.04 & -387.47 $\pm$ 0.04 & -419.50 $\pm$ 0.03 \\
J2004+1704 & 15 Sge B  & - & - & - & - &  \\
\hline
J2246+3319A$^{12}$ & HIP 112422 & 1890840149267988992 & 15.46 $\pm$ 0.81 & 159.06 $\pm$ 0.65 & 24.79 $\pm$ 0.80  \\
J2246+3319 & 2MASS J22461844+3319304  & - & - & - & - &  \\
\hline
J0219-3925A$^{13}$ & 2MASS J02192210-3925225 & 4963614887043956096 & 24.85 $\pm$ 0.09 & 103.70 $\pm$ 0.07 & -34.58 $\pm$ 0.08  \\
J0219-3925B & 2MASS J02192210-3925225B  & 4963614887043331072 & - & - & - &  \\
\hline
J1339+0104A$^{14}$ & HD 118865 & 3663438298389132416 & 16.50 $\pm$ 0.02 & -95.58 $\pm$ 0.03 & -48.19 $\pm$ 0.02  \\
J1339+0104 & HD 118865B  & - & - & - & - &  \\
\hline
J0429-3123$^{15}$ & 2MASS J04291842-3123568 & 4872659466967320576 & 58.79 $\pm$ 0.14 & 65.82 $\pm$ 0.15 & 99.58 $\pm$ 0.16  \\
J0429-3123B & 2MASSI J0429184-312356B  & - & - & - & - &  \\
\hline
J2005+5424A$^{16}$ & Wolf 1130 & 2185710338703934976 & 60.30 $\pm$ 0.03 & -1159.52 $\pm$ 0.04 & -904.01 $\pm$ 0.03 \\
J2005+5424B & Wolf 1130 B & - & - & - & - &  \\
J2005+5424 & Wolf 1130 C  & - & - & - & - &  \\
\hline
J0004-4044$^{17}$ & GJ 1001 A & 4996141155411983744 & 81.22 $\pm$ 0.03 & 677.68 $\pm$ 0.03 & -1505.62 $\pm$ 0.03 \\
J0004-4044B & GJ 1001 B & 4996141155411984128 & 82.35 $\pm$ 0.26 & 668.89 $\pm$ 0.23 & -1498.24 $\pm$ 0.20 &  \\
J0004-4044C & GJ 1001 C  & - & - & - & - &  \\
\hline
J0024-2708$^{18}$ & LHS 1070 & 2322561156529549440 & 129.32 $\pm$ 0.13 & -92.74 $\pm$ 0.11 & 695.96 $\pm$ 0.12\\
J0024-2708B & LHS 1070B  & - & - & - & - &  \\
J0024-2708C & LHS 1070C  & - & - & - & - &  \\
\hline
J0223+5240A$^{19}$ & HIP 11161 & 452046549154458880 & 12.65 $\pm$ 0.03 & -81.12 $\pm$ 0.03 & -61.06 $\pm$ 0.03  \\
J0223+5240B & Gaia DR3 452046549149685504 & 452046549149685504 & 12.44 $\pm$ 0.07 & -82.74 $\pm$ 0.09 & -55.14 $\pm$ 0.13 &  \\
J0223+5240 & 2MASS J02233667+5240066 & - & - & - & - &  \\
\hline
J1047+4026$^{20}$ & LP 213-67 & 779689606794219136 & 40.43 $\pm$ 0.07 & -298.29 $\pm$ 0.07 & -33.26 $\pm$ 0.07  \\
J1047+4027 & LP 213-68 & 779689533779300736 & 39.47 $\pm$ 0.33 & -301.66 $\pm$ 0.30 & -33.98 $\pm$ 0.33 &  \\
J1047+4027B & LP 213-68B & - & - & - & - &  \\
\hline
J1112+3548$^{21}$ &  GJ 417A & 761919883981626752 & 44.05 $\pm$ 0.02 & -249.19 $\pm$ 0.02 & -151.40 $\pm$ 0.03  \\
J1112+3548B & GJ 417B & 761918578311083264 & 42.24 $\pm$ 1.05 & -236.90 $\pm$ 0.88 & -149.79 $\pm$ 1.04 &  \\
J1112+3548C & GJ 417C & - & - & - & - &  \\
\hline
J1450+2354$^{22}$ & Gl 564 & 1265976524286377856 & 54.95 $\pm$ 0.03 & 144.40 $\pm$ 0.02 & 31.66 $\pm$ 0.04  \\
J1450+2354C & Gl 564 B & - & - & - & - &  \\
J1450+2354B & Gl 564 C & - & - & - & - &  \\
\hline
J2203-5647$^{23}$ & eps indi A & 6412595290592307840 & 274.84 $\pm$ 0.10 & 3996.66 $\pm$ 0.09 & -2536.19 $\pm$ 0.09   \\
J2204-5646B & eps Indi Ba & 6412596012146801152 & 270.66 $\pm$ 0.69 & 3981.98 $\pm$ 0.60 & -2466.83 $\pm$ 0.63 &  \\
J2204-5646C & eps Indi Bb & - & - & - & - &  \\ 
\hline
J2331-0405A$^{24}$ & HD 221356 & 2633449134434620288 & 38.71 $\pm$ 0.02 & 178.13 $\pm$ 0.03 & -191.84 $\pm$ 0.02  \\
J2331-0405 & ** GZA 1B & 2633449095780158720 & 38.54 $\pm$ 0.36 & 169.94 $\pm$ 0.40 & -194.59 $\pm$ 0.30 &  \\
J2331-0406 & 2MASSW J2331016-040618 & 2633426216489127296 & 38.51 $\pm$ 0.16 & 176.34 $\pm$ 0.17 & -189.38 $\pm$ 0.13 &  \\
J2331-0406B & 2MASSW J2331016-040618B & - & - & - & - &  \\ 
\bottomrule
\end{tabular}

\smallskip
\footnotesize{Binarity References : 
$^1$ - \cite{2005A&A...435L...5F}, 
$^2$ - \cite{2013ApJ...774...55B}, 
$^3$ - \cite{2014ApJ...794..143B(LUWE12)}, 
$^4$ - \cite{2003A&A...411..157M}, 
$^{5, 12, 19}$ - \cite{2014ApJ...792..119D(4)}, 
$^6$ - \cite{2015ApJ...804...96G}, 
$^7$ - \cite{2017A&A...605L...9C}, 
$^8$ - \cite{2012ApJ...757..100D}, 
$^9$ - \cite{1987BAAS...19.1128S}, 
$^{10}$ - \cite{2006MNRAS.368.1917L}, 
$^{11}$ - \cite{2002ApJ...571..519L}, 
$^{13}$ - \cite{2015ApJ...806..254A},
$^{14}$ - \cite{2013MNRAS.433..457B}
$^{15}$ - \cite{2005ApJ...621.1023S}
$^{16}$ - \cite{2013ApJ...777...36M}, 
$^{17}$ - \cite{1994A&A...291L..47L},
$^{18}$ - \cite{2012A&A...545A..85R},
$^{20}$ - \cite{2000MNRAS.311..385G}, 
$^{21}$ - \cite{2000AJ....120..447K}, 
$^{22}$ - \cite{2002ApJ...567L.133P},
$^{23}$ - \cite{2003A&A...398L..29S}, 
$^{24}$ - \cite{2012MNRAS.427.2457G}
}

\end{table*}

\begin{table*}
\caption{Objects identified as part of a companion system in the UCDC but not in the CNS5. }
\label{cns5_table}
\begin{tabular}{@{}llllll@{}}
\toprule
SN & DN &  ID & $\varpi$ & Companion SN & Projected Separation \\ 
& & & (mas) & & (AU) \\ \midrule

J2237+3922 & {G 216-7B} & 1908305165624662272 & 47.6 $\pm$ 0.14 & J2237+39221 & 71.1 \\
J1239+5515A & {HD 110463} & 1571411233756646656 & 44.01 $\pm$ 0.02 & J1239+5515 & 4630  \\
J0821+1443A & {2MASS J08213173+1443231} & 652005932802958976 & 43.9 $\pm$ 0.02 & J0821+1443 & 5470  \\
 \bottomrule
\end{tabular}
\end{table*}

\begin{table*}
\caption{Content of the UCDC catalogue. The associated errors are presented in the online version of this catalogue. This example shows a double system, but higher-multiplicity systems follow the same logic. The \texttt{ang\_sep},
\texttt{plx\_diff},
\texttt{pm\_diff} and 
\texttt{pma\_diff} are with respect to the faintest companion in the system.}
\begin{adjustbox}{width=1\textwidth}
\label{Table1}

\begin{tabular}{@{}lllll@{}}
\toprule
Parameter & Unit & Description & Star A & Star B \\ \midrule
\texttt{sys\_num} & ... & Number in which the system appears in the UCDC (this is the second system listed in the UCDC) & 2 & 2 \\

\texttt{num\_comp} & ... & Total number of objects that make up the system & 2 & 2 \\
\texttt{SN} & ... & Short Name - Unique name used in this catalogue & J0003-2822A & J0003-2822 \\
\texttt{DN} & ... & Discovery Name - Known identifier from literature & HD 225118 & 2MASS J00034227-2822410 \\
\texttt{Binary Discovery Ref} & ... & Binarity discovery bibcode  & -
 & 2021A\&A...649A...6G \\
\texttt{source\_id} & ... & Unique source identifier for Gaia DR3 & 2333120282087272192 & 2333120453885963392 \\
\texttt{FPP} & ... & False-Positive Probability - Probability of the system being a False-Positive match &  0.0001018 & -\\
\texttt{ang\_sep} & $\arcsec$ & Angular Separation between companion and UCD & 66.001 & - \\
\texttt{projected\_sep} & AU & Projected physical separation between companion and UCD & 2658 & - \\
\texttt{plx\_diff} & mas & Difference in Parallax between companion and UCD & 0.0802 & - \\
\texttt{pm\_diff} & mas yr$^{-1}$ & Proper Motion difference between companion and UCD & 4.685 & - \\
\texttt{pma\_diff} & deg & Proper Motion angle difference between companion and UCD & 0.0214 & -  \\
\texttt{ra} & deg & Right Ascension & 0.9242 & 0.9277 \\
\texttt{dec} & deg & Declination & -28.40 & -28.39 \\
\texttt{epoch} & ...& Epoch of object position & 2016.5 & 2016.5\\
\texttt{epoch\_ref} & ...& Bibcode for Epoch position & 2021A\&A...649A...1G & 2021A\&A...649A...1G\\
\texttt{parallax} & mas & Parallax & 24.83 & 24.75 \\
\texttt{parallax\_ref} & ... & Bibcode for parallax value & 2021A$\&$A...649A...1G & 2021A$\&$A...649A...1G \\
\texttt{pmra} & mas yr$^{-1}$ & Proper Motion in Right Ascension direction & 281.1 & 285.4 \\
\texttt{pmdec} & mas yr$^{-1}$ & Proper Motion in Declination direction & -142.1 & -144.1 \\
\texttt{pm} & mas yr$^{-1}$ & Total Proper Motion & 315.0 & 319.7 \\
\texttt{pm\_ref} & ... & Bibcode for Proper Motion value & 2018A\&A...616A...1G & 2018A\&A...616A...1G \\
\texttt{G} & mag & G-band mean magnitude & 8.084 & 17.33 \\
\texttt{G$\rm{_{rp}}$} & mag &  RP mean magnitude & 7.537 & 15.78 \\
\texttt{G-\texttt{G$\rm{_{rp}}$}} & mag & \texttt{G-\texttt{G$\rm{_{rp}}$}} colour & 0.55 & 1.55 \\
\texttt{J} & mag & 2MASS J-band & 6.970 & 13.07 \\
\texttt{H} & mag & 2MASS H-band & 6.621 & 12.38 \\
\texttt{K$\rm{_s}$} & mag & 2MASS \texttt{K$\rm{_s}$}-band & 6.554 & 11.97 \\
\texttt{sptopt} & ... & Optical Spectral Type & G8.5V & M8.0 \\
\texttt{sptopt\_ref} & ... & Optical Spectral Type reference & 2006AJ....132..161G & 2014AJ....147..160M \\
\texttt{sptnir} & ... & Near Infrared Spectral Type & - & - \\
\texttt{sptnir\_ref} & ... & Near Infrared Spectral Type reference & - & - \\
\texttt{astrometric\_n\_obs\_al} & ... & Total number of observations in the along-scan (AL) direction & 244 & 258 \\
\texttt{phot\_g\_n\_obs} & ... & Number of observations contributing to G photometry & 284 & 300 \\
\texttt{phot\_rp\_n\_obs} & ... & Number of observations contributing to RP photometry & 33 & 25 \\
\texttt{phot\_rp\_n\_blended\_transits} & ... & Number of RP blended transits & 0 & 0 \\
\texttt{ruwe} & ... & Renormalised Unit Weight Error & 0.9553 & 1.141 \\
\texttt{luwe} & ... & Local Unit Weight Error & - & - \\
\texttt{ipd\_gof\_harmonic\_amplitude} & deg & Amplitude of the IPD GoF versus position angle of scan & 0.0298 & 0.0042 \\
\texttt{ipd\_frac\_multi\_peak} & $\%$ & Percent of successful-IPD windows with more than one peak & 0 & 0 \\
\texttt{rv} & km s$^{-1}$ & Radial Velocity & 10.90 & - \\
\texttt{rv\_ref} & ... & Bibcode for Radial Velocity value  & 2011AJ....141...97W & - \\ 
\texttt{v$_{\rm TAN}$} & km s$^{-1}$ & Tangential Velocity & 60.13 & 61.23\\
\texttt{T$\rm{_{eff}}$} & K & Effective temperature  & 5398 & 2475 \\
\texttt{T$\rm{_{eff}}$\_ref} & ... & 1 = GSP-Phot, 2 = ESP-UCD & 1 & 2 \\
\texttt{logg\_gspphot} & log(cgs) & Surface gravity from GSP-Phot (DR3) (Median of MCMC values) & 4.408 & - \\
\texttt{mh\_gspphot} & dex & Iron abundance from GSP-Phot (DR3) (Median of MCMC values) & 0.1329 & - \\ 
\texttt{$\log(L/L_{\odot})$} & ... & Bolometric luminosity for objects discussed in Sec \ref{mass_age_section}  & - & - \\ 
\texttt{mass} & \(M_\odot\) & Mass of the object & 0.86 & - \\
\texttt{mass\_ref} & ... & Mass reference, 1 = $\gaia$ FLAME, 2 = White Dwarf, 3 = BANYAN $\Sigma$ & 1 & - \\
\texttt{age} & Gyr & Age of the object (always assumed to be coeval) & 13.32 & 13.32\\
\texttt{age\_ref} & ... & Age reference, 1 = $\gaia$ FLAME, 2 = White Dwarf, 3 = BANYAN $\Sigma$ & 1 & 1 \\
 
\bottomrule
\smallskip
\end{tabular}
\end{adjustbox}
\footnotesize{ \scriptsize
{UCD spectral type references: \\
\cite{2020AJ....159..257B}, \cite{1999ApJ...519..802K}, \cite{2019AJ....157..231K}, \cite{2019MNRAS.486.1840Z}, \cite{2008ApJ...689.1295K}, \cite{2000AJ....120..447K}, \cite{2008AJ....135..785W}, \cite{2001AJ....121.2185G}, \cite{2001AJ....121.3235K}, \cite{2010ApJS..190..100K}, \cite{2011AJ....141...97W}, \cite{2000MNRAS.311..385G}, \cite{2015ApJ...804...96G}, \cite{2008MNRAS.388..838D}, \cite{2016ApJ...817..112S}, \cite{2014AJ....147..160M}, \cite{2006ApJ...645L.153P}, \cite{2010ApJ...710...45B}, \cite{2000A&A...360L..39N}, \cite{2000MNRAS.311..456M}, \cite{2012ApJ...752...56F}, \cite{2015ApJS..216....7B}, \cite{2007AJ....134.2340K}, \cite{2017A&A...605L...9C}, \cite{2018AJ....156...76L}, \cite{2007ApJ...667..520C}, \cite{2003AJ....126.2487B}, \cite{2015ApJ...802...37B}, \cite{2019AJ....158..182G}, \cite{2001AJ....122.1989W}, \cite{2002AJ....123.3409H}, \cite{2014MNRAS.445.3908L}, \cite{2016ApJS..224...36K}, \cite{2000ApJ...538..363B}, \cite{2015ApJS..219...33G}, \cite{2013A&A...557L...8B}, \cite{2017MNRAS.465.4723K}, \cite{2010AJ....139.1808S}, \cite{1973MNRAS.163..381W}, \cite{2014ApJ...781L..24S}, \cite{2015A&A...577A.128A}, \cite{2010MNRAS.404.1817Z}, \cite{2000AJ....120.1085G}, \cite{2003AJ....126.2421C}, \cite{2016ApJS..225...10F}, \cite{2007AJ....133.2258S}, \cite{2010AJ....139.2566D}, \cite{2002ApJ...575..484G}, \cite{1990ApJ...354L..29H}, \cite{2008MNRAS.383..831P}, \cite{2011ApJ...743..138G}, \cite{2008AJ....136.1290R}, \cite{2007AJ....133..439C}, \cite{2014MNRAS.441.2644C}, \cite{2017MNRAS.470.4885M}, \cite{1995AJ....109..797K}, \cite{2009AJ....137.3345C}, \cite{2006A&A...459..511B}, \cite{2015ApJ...814..118B}, \cite{2007MNRAS.378.1328M}, \cite{2013ApJS..205....6M}, \cite{2017ApJS..231...15D}, \cite{2015AJ....149..104B}, \cite{2014ApJ...794..143B}, \cite{2021ApJS..253....7K}, \cite{2002ApJ...567L..59G}, \cite{2010AJ....139.2448B}, \cite{2014AJ....147...20N}, \cite{2006ApJ...639.1095B}, \cite{2013AJ....146..161M}, \cite{2002ApJ...568L.107L}, \cite{2015MNRAS.454.4476S}, \cite{2012ApJ...748...74L}, \cite{2011ApJ...729..139W}, \cite{2011ApJS..197...19K},  \cite{2020MNRAS.494.4891M}, \cite{2018AJ....156...57D}, \cite{2006AJ....131.2722C}, \cite{2012ApJS..201...19D}, \cite{2014ApJ...787..126L}, \cite{2006AJ....131.1007B},  \cite{2008ApJ...689..471R}, \cite{2004AJ....127.3553K}, \cite{2015ApJ...806..254A}, \cite{2016AJ....151...46A}, \cite{2006ApJ...651.1166M}, \cite{2013MNRAS.431.2745G}, \cite{2014AJ....147...34S}, \cite{2013ApJ...774...55B}, \cite{2009AJ....138.1563B}, \cite{2010ApJ...710.1142B}, \cite{2013MNRAS.433..457B}, \cite{2011ApJ...739...81L}, \cite{2021ApJ...920...20C}, \cite{2007A&A...466.1059K}, \cite{2011MNRAS.414..575M},  \cite{2015ApJ...804...92S}, \cite{2012ApJ...757..110B}, \cite{2012MNRAS.427.2457G}, \cite{2015MNRAS.449.3651M}, \cite{2012ApJ...755...94D}, \cite{2004A&A...427L...1F}, \cite{2000ApJ...529L..37M}, \cite{2009AJ....137....1F}, \cite{2009MNRAS.395.1237B},  \cite{2016ApJ...830..144R}, \cite{2004ApJ...617.1330M}, \cite{2013PASP..125..809T}, \cite{2012ApJ...760..152L}, \cite{2014ApJ...792..119D}, \cite{2012A&A...545A..85R}, \cite{2012ApJ...757..100D}, \cite{2012MNRAS.422.1922P}, \cite{2013ApJ...772...79A}, \cite{2007ApJ...654..570L} \\
\newpage
Binary discovery references: \\
\cite{2012ApJ...757..110B}, \cite{2007A&A...462L..61C}, \cite{2005A&A...435L...5F}, \cite{2006MNRAS.373L..31M}, \cite{2013ApJ...774...55B}, \cite{2013MNRAS.434..142B}, \cite{2001AJ....121.2185G}, \cite{2018AJ....156...57D}, \cite{2006ApJ...645L.153P}, \cite{2005ApJ...621.1023S}, \cite{2012AJ....144..180M}, \cite{2011ApJ...729..139W}, \cite{1995Natur.378..463N}, \cite{2005A&A...438L..29C}, \cite{2011ApJ...739...81L}, \cite{2004ApJ...617.1330M}, \cite{2007ApJ...667..520C}, \cite{2021ApJ...916L..11Z}, \cite{2007MNRAS.378.1328M}, \cite{2012ApJ...755...94D}, \cite{2011ApJ...732L..29R}, \cite{2018ApJS..234....1B}, \cite{2014A&A...568A...6Z}, \cite{2005AN....326..974S}, \cite{2005A&A...430L..49S}, \cite{2021ApJS..257...45H}, \cite{2003ApJ...584..453F}, \cite{2011ApJS..197...19K}, \cite{1999ApJ...512L..69L}, \cite{2003AJ....125.3302G}, \cite{2016ApJ...818L..12S}, \cite{2019MNRAS.486.1840Z}, \cite{2010A&A...515A..92S}, \cite{2011AJ....141...70B}, \cite{2023AJ....166..103S}, \cite{2017A&A...605L...9C}, \cite{2013MNRAS.431.2745G}, \cite{2022ApJ...936..109P}, \cite{2013MNRAS.433..457B}, \cite{2008MNRAS.383..831P}, \cite{2012ApJ...757..100D}, \cite{2004ApJ...601..289C}, \cite{2010A&A...510L...8S}, \cite{2012MNRAS.422.1922P}, \cite{1992ApJ...386..260Z}, \cite{2004A&A...427L...1F}, \cite{2000ApJ...529L..37M}, \cite{2002AJ....123.2806R}, \cite{2000ApJ...531L..57B}, \cite{2011MNRAS.414..575M}, \cite{2002ApJ...575..484G}, \cite{2015MNRAS.454.4476S}, \cite{2007ApJ...658..557B}, \cite{2009AJ....138.1563B}, \cite{2014ApJ...787..126L}, \cite{2008ApJ...689..471R}, \cite{2011ApJ...743..109S}, \cite{2013A&A...557L...8B}, \cite{2006MNRAS.368.1917L}, \cite{2006A&A...460L..19M}, \cite{2002ApJ...571..519L}, \cite{2007AJ....134.1162L}, \cite{2009ApJ...707L.123T}, \cite{2003A&A...401..677G}, \cite{2006AJ....131.1007B}, \cite{2021RNAAS...5..129C}, \cite{2001PASP..113..814K}, \cite{2010ApJ...715..561A}, \cite{2004AJ....128.1733G}, \cite{1994A&A...291L..47L}, \cite{2023A&A...674A..39G}, \cite{2014MNRAS.441.2644C}, \cite{2014MNRAS.445.3908L}, \cite{2017MNRAS.466.2983G}, \cite{2006PASP..118..671R}, \cite{2001AJ....121.3235K}, \cite{2003ApJ...587..407C}, \cite{2003AJ....126.1526B}, \cite{2001AJ....122.3466M}, \cite{2021A&A...649A...6G}, \cite{2002ApJ...567L.133P}, \cite{2013ApJ...777...36M}, \cite{2018A&A...619A..32B}, \cite{2006ApJ...651.1166M}, \cite{2004MNRAS.347..685S}, \cite{2016MNRAS.457.3191D}, \cite{2012MNRAS.427.2457G}, \cite{2003A&A...398L..29S}.

}
}
\end{table*}

\section{Identification of Unresolved UCD Binaries}
\label{indentify binaries}
\subsection{Finding Binaries From Astrometric Solutions}
\label{LUWE_section}
\cite{2022MNRAS.513.5270P} presents a catalogue of unresolved candidate binary systems in the GCNS identified from astrometric deviations alone. This study uses the Renormalised Unit Weight Error (RUWE), which is a rescaling of the Unit Weight Error (UWE) from the 5-parameter astrometric solution, to account for the astrometric error. The UWE is calculated using the entire $\gaia$ catalogue and is dominated by objects beyond the 100 pc limit of the GCNS. This results in distant objects having an inordinate weighting of the RUWE due to the volume of objects of similar colour further than 100 pc. To correct for this bias, \cite{2022MNRAS.513.5270P} rescaled the RUWE by finding a correction factor based on the apparent magnitude and colour of each source over a uniform sample (GCNS), resulting in the Local Unit Weight Error (LUWE). A well-behaved single star is expected to have a LUWE value of approximately 1.0, and any value significantly larger than this may indicate the possibility of unresolved binarity. Specifically, unresolved binarity is defined as LUWE $>$ 2 and $\Delta$LUWE $>$ - $\frac{\text{LUWE}}{3}$, where $\Delta$LUWE is the difference in LUWE between DR2 and EDR3.

In this study, as in \cite{2022MNRAS.513.5270P}, UCDC systems have been identified through astrometry alone. 
The cleaning of spurious sources in the GCNS and strict binarity criteria imposed by \cite{2022MNRAS.513.5270P} results in only 1$\%$ of the catalogue being occupied by the low-mass regime, which significantly reduces the number of potential systems for analysis. However, despite this numerical limitation, the high-resolution and precise measurements provided by the GCNS and LUWE datasets not only help to confirm the existing companion candidates identified through astrometry but also allow for the potential discovery of new, previously overlooked systems. \newline
The catalogue produced by \cite{2022MNRAS.513.5270P} contains 22,699 objects with $\gaia$ source IDs, which were cross-matched with the UCDC, resulting in 30 matched objects, listed in Table \ref{LUWE_table} with their respective companions. 
The LUWE values are smaller in most cases, indicating overestimated RUWE values for these objects. Notably, a RUWE $\ge$ 1.4 is indicative of a Non-Single Source (NSS) and thus a poor astrometric solution, of which 110 objects in the UCDC exceed.

From the matched systems there are four with $\rho$ $<$ 0.18 $\arcsec$ which is smaller than the $\gaia$ limit for distinguishing between individual sources. Any separation below this is considered a duplicate and removed from the $\gaia$ catalogue \cite[there are some exceptions; see ][]{EDR3_astrometry} indicating that these objects are in companion systems with non-$\gaia$ objects. Given that the separations are smaller than the $\gaia$ limit, it can be assumed that the companions are likely responsible for the observed astrometric deviations. \\
To explore the likelihood of hidden binarity, we utilise two measurements from $\gaia$: \textbf{ipd\_gof\_harmonic\_amplitude} and \textbf{ipd\_frac\_multi\_peak}. The former measures the significance of the scanning angle in PSF fitting indicating elongation and the presence of partially resolved objects. The latter quantifies the percentage of scans in which two peaks are visible in the image, thereby revealing the potential presence of a secondary source. 
The Image Parameter Determination (IPD) quantification is most significant below 2$\arcsec$ \cite[as displayed in Fig.18 of ][]{2021MNRAS.506.2269E}, but is limited to 0.18$\arcsec$, meaning that the IPD measurements are sensitive between this range, both measurements are included in Table \ref{LUWE_table}.\\
The UCDC omits sources in $\gaia$ with 2-parameter solutions due to the lack of parallax and proper motion as just the positional data is given for these sources. If such a UCD is in a close companion system, it will not be identified because of the selection criteria explained in Sec.\ref{adapted_cuts}; however, assuming that the primary has a full solution, the IPD flags and RUWE should give an indication of an unresolved companion, provided a notable astrometric deviation is measured. 

The objects listed in Table \ref{LUWE_table} are shown in Fig. \ref{fig:LUWE}. Fig. \ref{fig:3a} exhibits remarkably red UCDs which are difficult to explain by intrinsic properties such as age, atmosphere, or composition alone. This suggests that the $\grp$  values for these UCDs may be inherently biased due to possible blending effects (see Sec \ref{blended section} for further detail) from the unresolved companion.
The longer baseline of the $\g$–K$\rm{_s}$ colour should provide a more robust representation of the underlying companion systems among the UCDs. This is evident as the UCDs in Fig.\ref{fig:3b} appear to be residing on the binary Main-Sequence, with most of these systems located above the MS, including the non-UCD object, an expected photometric signature of unresolved binarity.

\subsubsection{Notable Systems found from LUWE}
We can immediately judge the usefulness of the LUWE as 4/7 UCDs in Table \ref{LUWE_table} have been studied in previous literature, identifying close low-mass companions that are unresolvable to $\gaia$, including \textbf{J1047+4027} \citep{2003ApJ...587..407C}, \textbf{J1121-1313} \citep{2003A&A...411..157M}, \textbf{J0429-3123} \citep{2005ApJ...621.1023S}, and \textbf{J0021-4244} \citep{2006AJ....132..663B}. Studying a sample of the non-UCD targets in Table \ref{LUWE_table} also reveals known close companions including J1559+4403 \citep{2015ApJ...806...62B}, J2126-8140A\citep{2016PASP..128j2001B} and J1712-0507A
\citep{2011ApJ...743..109S}. As these systems are known to be close companions, their identification through LUWE validates this approach as an effective measure for detecting binarity, particularly in the realm of low-mass, closely bound systems that typically elude resolution by $\gaia$. The remaining 3 UCDs in the LUWE sample are discussed below: 

(i) \textbf{J2325+4608A}  lacks a SIMBAD entry; however, it was studied in \cite{2020MNRAS.494.4891M(luwe20)}, as they also noticed a significant overluminosity ($\sim$ 2.5\,mag) above the MS. They concluded that this was not a consequence of an equal-mass binary, as this would result in only 0.75\,mag elevation above the MS and the redness cannot be indicative of youth, as it is not known in any young moving groups known in BANYAN $\Sigma$. The possibility of contamination from a background object was also discussed, but disregarded, as this is an improbable scenario. J2325+4608A shows no signs of peculiarity in its photometry in 2MASS, ALLWISE, or PS1, but all absolute magnitudes are overluminous. \cite{2020MNRAS.494.4891M(luwe20)} conducted the study using $\gaia$ DR2 data which for this object lacked IPD values which indicate a strong possibility of binarity (\textbf{ipd\_gof\_harmonic\_amplitude} = 0.126,  \textbf{ipd\_frac\_multi\_peak} = 70$\%$) which cannot be attributed to its wide companion due to the large angular separation between them ($\rho$ = 7.2$\arcsec$). The parallax also differs from $\varpi_{DR3}$ = 21.53 $\pm$ 0.29\,mas and $\varpi_{DR2}$ = 19.13 $\pm$ 0.48\,mas, resulting in the object appearing redder from DR3 photometry $\g$-$\grp$ = 1.76\,mag compared to 1.59\,mag of DR2 and fainter in DR3 M$_{\rm{G}}$ = 13.28\,mag compared to M$_{\rm{G}}$ = 12.84\,mag in DR2 but still too overluminous to suggest an equal-mass companion. UKIRT UHS-J band \citep{2018MNRAS.473.5113D} imaging is available but shows no discernible elongation of the PSF. Given the significant redness, IPD flags, and poor single-solution astrometric fit indicated by a large LUWE, we propose that this is most likely an unresolved binary with an underestimated parallax uncertainty contributing to the observed significant overluminosity. 

\begin{figure*}
  \centering
  \begin{subfigure}{0.5\textwidth}
    \centering
    \includegraphics[clip,width=\linewidth]{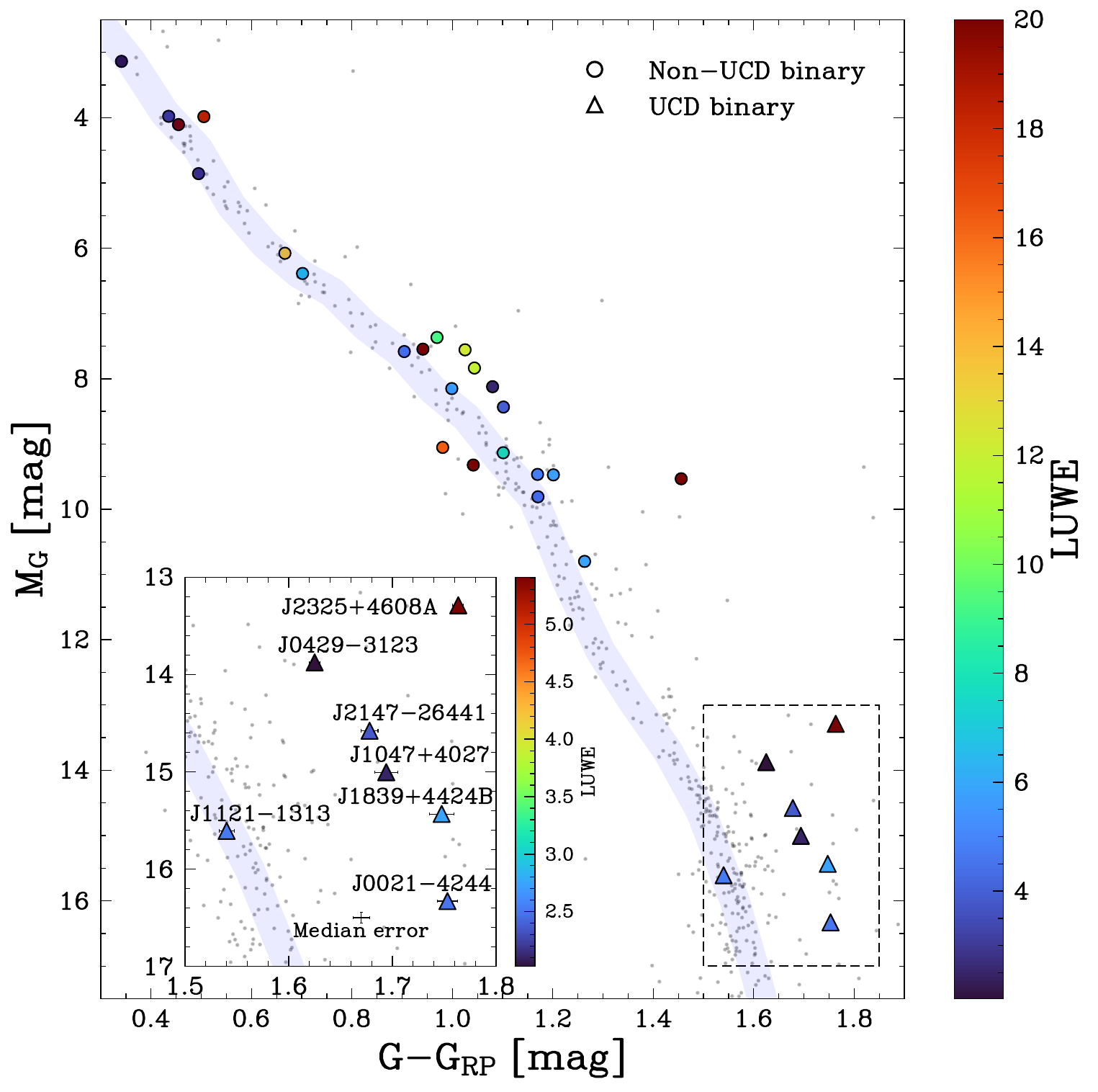}
    \caption{}
    \label{fig:3a}
  \end{subfigure}%
  \begin{subfigure}{0.5\textwidth}
    \centering
    \includegraphics[clip,width=\linewidth]{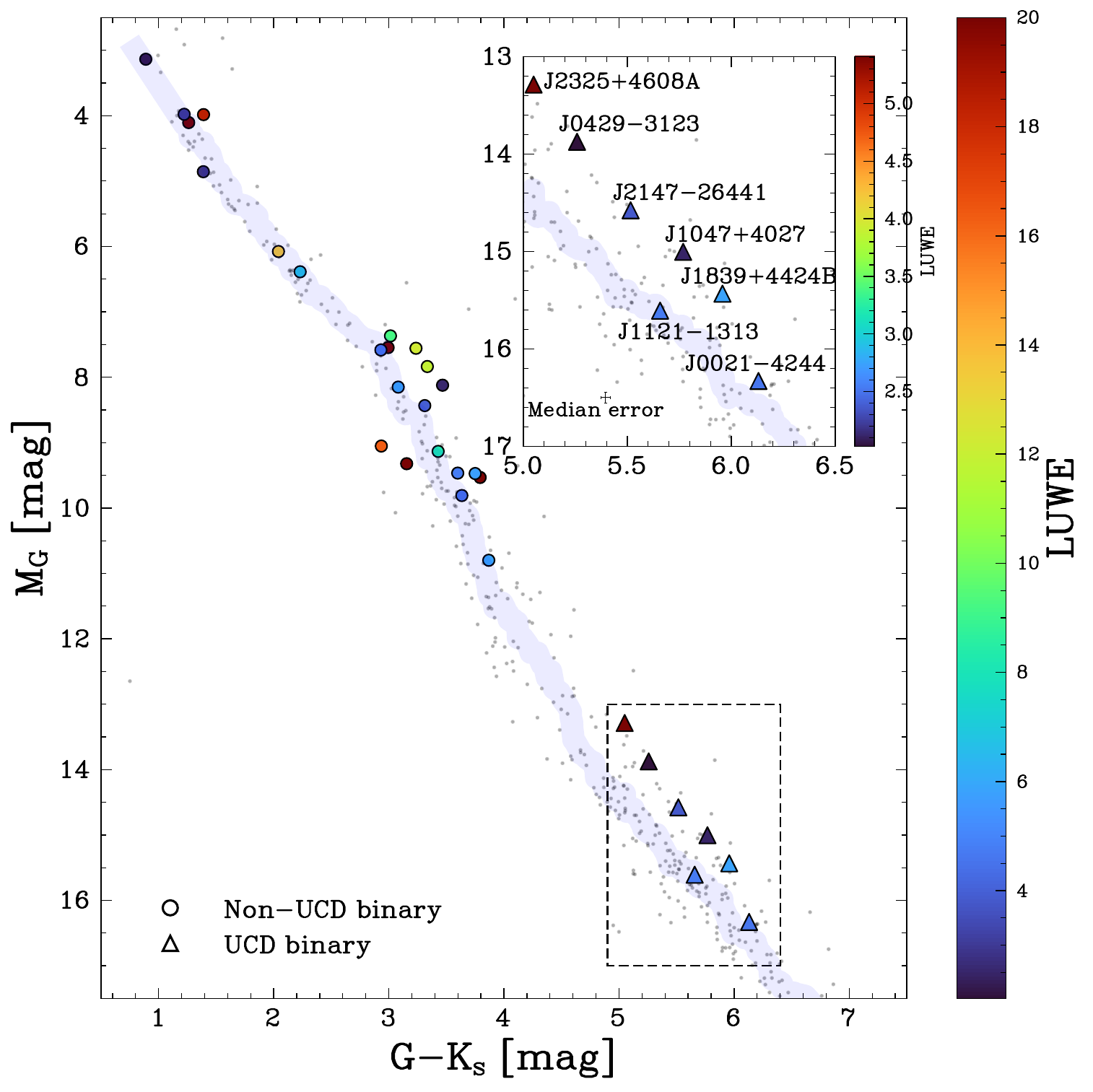}
    \caption{}
    \label{fig:3b}
  \end{subfigure}
  
  \caption{CMD of sources that exceed the LUWE criteria, as listed in Table \ref{LUWE_table}, are depicted for two different colour indices: $\g$-$\grp$ on the left (3a) and $\g$-K$\rm{_s}$ on the right (3b). The regions encompassed by the black dashed lines represent the UCD sample which the subplots focus on. The colour of the marker represents the LUWE of each star.}
  \label{fig:LUWE}
\end{figure*}

(ii) \textbf{J1839+4424B} is an M9 \citep{2003AJ....126.2421C} with a BD candidate companion J1839+4424A \citep{2018A&A...619L...8R}, separated by $\rho$ = 21.9$\arcsec$. From Fig. \ref{fig:3b}, the object resides $\sim$ 0.75\,mag above the MS making the position on the CMD consistent with a near-equal mass binary companion. The IPD flags suggest a degree of non-isotropic structure, which can be seen in the UHS-J band, with a slight elongation observed to the southwest. 
A nearby $\gaia$ source (ID = 2117179153330392320), positioned 1.8$\arcsec$ away from J1839+4424B in the sky, has a parallax of $\varpi$ = 2.91 $\pm$ 0.12\,mas. Both objects are faint ($\g$ $>$ 18\,mag), suggesting an inherent bias may exist with the IPD flags as the images are collapsed to 1D (for objects with $\g$ $>$ 13\,mag) with similar brightness and close separation, which may explain the problematic astrometric solutions observed.

(iii) \textbf{J2147-26441} resides $\sim$\ 0.75\,mag above the MS in Fig. \ref{fig:3b} suggesting a near-equal mass binary companion. Despite no known $\gaia$ source existing within $\sim$ 1$\arcsec$, there are blended transits for $\grp$ and $\gbp$ (1/47 and 2/41 transitions, respectively), suggesting that a companion is on the fringe of what is resolvable to $\gaia$ and is scan-angle-dependent, which is corroborated by its large \textbf{ipd\_gof\_harmonic\_amplitude} = 0.15, with some elongation in the PSF predominately in the south in \textit{i}-band of PS1. This also explains the unusual redness in $\g$-$\grp$ = 1.68 which is significantly higher than that expected for a typical M7 \cite[$\g$ - $\grp$ = 1.46, ][]{2019AJ....157..231K}. Intriguingly, J2147-26441 has been catalogued in the Gaia Nearby Accelerating Star Catalogue (GNASC) \citep{2023AJ....165..193W}. The GNASC and implications of proper motion anomalies (PMa) for the identification of UCD binaries are further discussed in Sec.\ref{NSS_section}. J2147-26441 is found to be accelerating, displaying a notable (5$\sigma$ confidence) PMa. This could be indicative of a gravitationally bound companion affecting the motion of J2147-26441, hence hinting at binarity.

\begin{table*}
\centering
\caption{Objects in the UCDC which exceeds the LUWE \citep{2022MNRAS.513.5270P} criteria for a single source listed with their respective companions. \textbf{RUWE} - Renormalised Unit Weight Error, \textbf{LUWE} - Local Unit Weight Error, \textbf{amplitude} - ipd\_gof\_harmonic\_amplitude, 
\textbf{multi\_peak - ipd\_frac\_multi\_peak}.}
\label{LUWE_table}
\begin{tabular}{@{}llllrlllll@{}}
\toprule
SN & ID & RUWE & LUWE & $\rho$ & SpT & amplitude & multi\_peak & Companion SN  & Companion SpT \\
& & & & ($\arcsec$) & &  & ($\%$) & \\  \midrule
J0021-4244 & 4992141475707360640 & 2.54 & 2.50 & 77.8 & M9V  & 0.16 & 0 & J0021-4245 & M6  \\
J0025+4759A & 392562179817077120 & 18.4 & 18.4 & 217.5 & F8 &  0.023 & 22 & J0025+4759 & L4 \\
J0122+0331A & 2562742595373569280 & 2.83 & 2.86 & 44.8 & G5V  & 0.073 & 1 & J0122+0331 & L1 \\
J0219+1943 & 87135785400708992 & 9.41 & 8.19 & 1120 & -  & 0.050 & 0 & J0221+1940 & M8  \\
J0429-3123 & 4872659466967320576 & 5.07 & 2.02 & 0.06 & M7V  & 0.059 & 79 & J0429-3123B & - \\
J0805+5109 & 935430721585996800 & 46.0 & 34.9 & 0.16 & -  & 0.17 & 77 & J0805+5113 & L1  \\
J0840+2313A & 666141254289307264 & 5.41 & 4.80 & 351.6 & M3  & 0.015 & 0 & J0840+2313 & M7 \\
J0850-03181 & 5762038930729097728 & 6.77 & 5.88 & 8.84 & M1 & 0.029 & 36 & J0850-0318 & M8V  \\
J0900+3205A & 712049433864692352 & 2.19 & 2.23 & 35.3 & F2  & 0.031 & 0 & J0900+3205 & L1V  \\
J0907+4509B & 1009609617150473600 & 6.94 & 6.70 & 301.1 & -  & 0.017 & 0 & J0907+4509 & L0  \\
J0933-2752A & 5633969259436187648 & 4.82 & 3.92 & 29.3 & -  & 0.001 & 1 & J0933-2752 & M7V  \\
J1043-17061 & 3557078484185457280 & 6.70 & 5.80 & 17.02 & M4  & 0.029 & 0 & J1043-1706 & M9  \\
J1121-1313 & 3562157781229213312 & 3.12 & 2.53 & 0.03 & M8V  & 0.037 & 0 & J1121-1313B & L7V  \\
J1217+1427A & 3921176983720146560 & 3.04 & 3.07 & 38.1 & F8  & 0.011 & 0 & J1217+1427 & L1 \\
J1238+6219A & 1583589939940520960 & 6.29 & 4.55 & 44.6 & -  & 0.006 & 2 & J1238+6219 & L0  \\
J1316+5735A & 1566379112632896384 & 6.87 & 5.84 & 43.04 & M3  & 0.027 & 0 & J1316+5735 & M9  \\
J1419+2041 & 1252184211771920128 & 13.7 & 13.6 & 57.9 & - & 0.028 & 0 & J1419+2042 & M7V  \\
J1449-0117A & 3650384995128745472 & 10.8 & 9.22 & 90.70 & M1e  & 0.046 & 0 & J1449-0117 & M8  \\
J1559+4403 & 1384769474242031616 & 13.9 & 11.9 & 5.62 & M2.0Ve &  0.004 & 0 & J1559+4404 & M8 \\
J1606+2253A & 1206502600310698880 & 21.0 & 21.2 & 35.3 & -  & 0.010 & 0 & J1606+2253 & M8 \\
J1712-0507A & 4364702279101280256 & 18.6 & 16.3 & 5.92 & G0 &  0.034 & 0 & J1712-0507 & sdM7\\
J1743+85264 & 1724494760222303872 & 24.4 & 20.8 & 29.7 & M2.0V & 0.016 & 0 & J1743+8526 & L5 \\
J1839+4424B & 2117179153332367232 & 2.80 & 2.74 & 21.9 & M9 &  0.22 & 0 & J1839+4424A & L2  \\
J2147-26441 & 6810425909217925760 & 2.40 & 2.36 & 214.8 & M7Ve  & 0.15 & 0 & J2147-2644 & - \\
J2325+4608A & 1938529473261040640 & 5.56 & 5.41 & 7.24 & M8  & 0.13 & 70 & J2325+4608B & L2 \\
J0856+3746A & 719505638825948928 & 15.2 & 12.2 & 891.1 & - &  0.011 & 0 & J0856+3746 & M8  \\
J1047+4027 & 779689533779300736 & 2.78 & 2.12 & 0.16 & M8 &  0.11 & 0 & J1047+4027B & L0  \\
J1627+33283 & 1325653136360249856 & 27.6 & 25.0 & 9.26 & K7V &  0.016 & 0 & J1627+3328 & M9  \\
J1924-6826B & 6421889221867428096 & 5.20 & 4.58 & 1500 & - &  0.007 & 5 & J1924-6826 & M9  \\
J2126-8140A & 6348514275456913792 & 2.91 & 2.54 & 217.5 & M1Ve & 0.051 & 24 & J2126-8140 & L3 \\ \bottomrule
\end{tabular}
\end{table*}

\subsection{ Binarity From Blended Photometry }
\label{blended section}

The $\gaia$ mission utilises a 3.5$\arcsec$ $\times$ 2.1$\arcsec$ window to extract integrated mean fluxes for both the $\gbp$ and $\grp$ bands \citep{grp_excess}. However, in situations involving sources nearby to each other, such as crowded fields or binary systems, the integrated mean flux may be overestimated because of blending effects, where the nearby secondary source contributes to the measured flux of the primary source. The effects of blending diminish as the difference in brightness between the sources increases, as the CCD window is generally assigned to the brighter source. The $\g$-band is determined from profile fitting to a 2-dimensional PSF for $\g \leq$ 13\,mag and a 1-dimensional Line Spread Function (LSF) otherwise \citep{2016A&A...595A...1G}. The dissimilarity in photometric measurement techniques results in the $\g$-band being less prone to bias from variations in the background level or proximate sources. As for DR3, there are no provisions for managing multiple sources within the same window \citep{2023A&A...674A...2D}.

\begin{figure*}
\begin{center}
\includegraphics[clip,width=2\columnwidth]{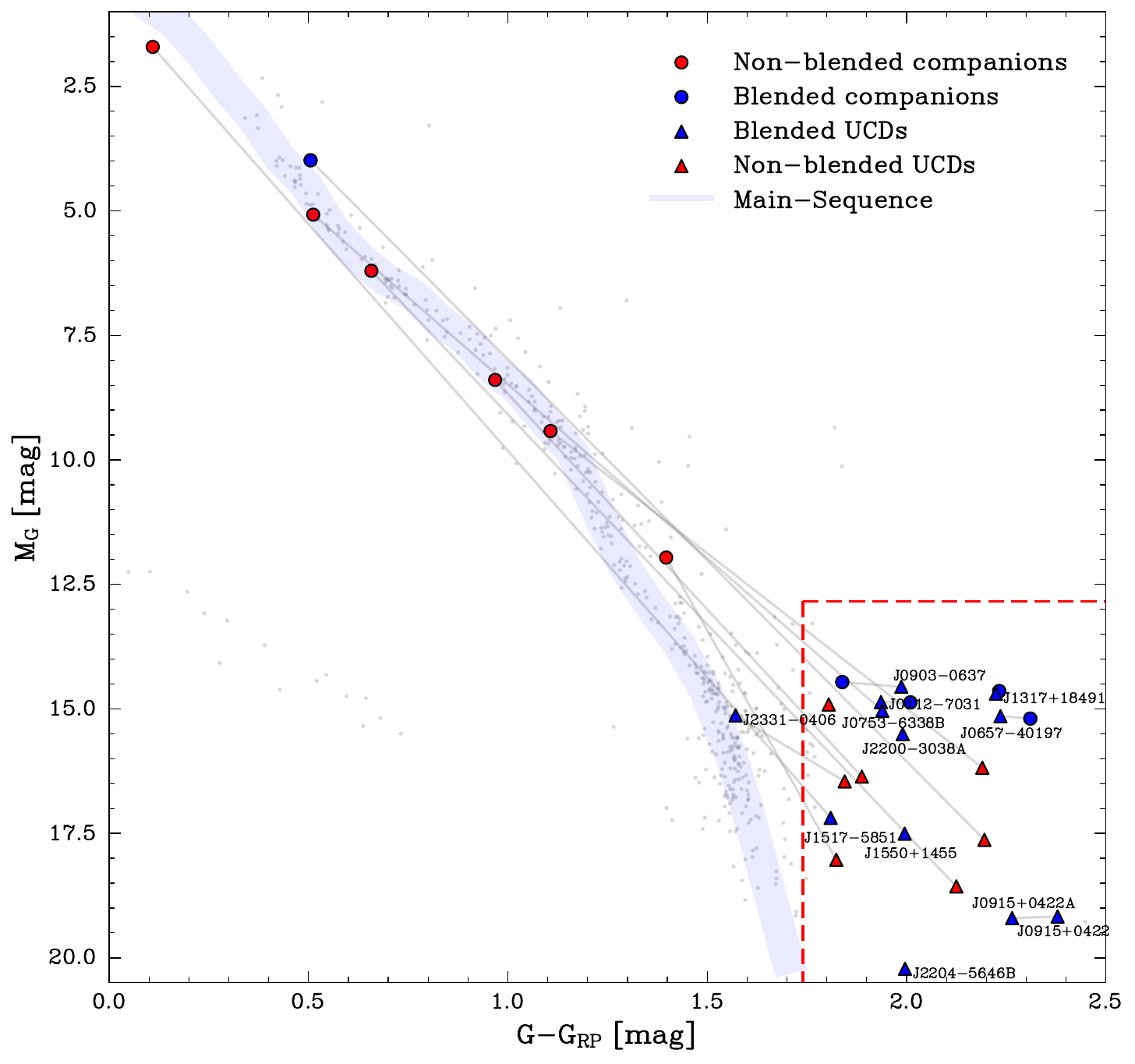}
\caption{Systems that include at least one significantly red UCD, as discussed in Sec.\ref{blended section} and detailed in Table \ref{blended_table}. Objects depicted with triangular markers correspond to spectroscopically confirmed UCDs, while those marked with circles are either non-UCDs or have yet to undergo spectroscopic follow-up to confirm their UCD status. Blue markers pass our criterion for blending ($\grp$ blended / $\grp$ obs $\ge$ 0.5), and red-coloured markers are not deemed to be blended. Companions within a system are interconnected by grey lines. The region enclosed by the red-dashed lines is explained in Sec.\ref{blended section}. Names are provided for UCDs that have been classified as blended. All objects from the  UCDC are overlaid in grey. Isolated objects with no connecting lines represent systems where the companion lacks a published $\grp$ value, and consequently, cannot be visually represented in this figure.}
\label{blended}
\end{center}
\end{figure*}

Blending effects on UCDs may lead to an overestimation of the $\grp$ flux because these sources emit primarily in the Near Infrared (NIR) region. Using the $\g$ - $\grp$ colour yields a relatively reliable $\g$ flux and an overestimated $\grp$ flux. Consequently, blended sources appear significantly redder on a CMD than well-modelled single stars \cite[further elaborated in][]{2023A&A...670A..19G}. A targeted approach is necessary to systematically investigate the effects of blending. One useful strategy involves defining a specific region in the CMD phase space, where blending impacts can be most readily identified. Fig. \ref{fig:cmd} highlights the region where blending effects are likely to be observed, denoted as "Unresolved Binaries". We concentrated on the reddest objects in our sample, primarily because the blending effects become markedly pronounced among such objects. By constraining the phase space of the CMD, we established a limiting magnitude of M$_\g$ $\ge$ 12.84\,mag. This decision was based on empirical evidence from one of the brightest identified M7 objects in the Rho Ophiuchi star cluster, UScoCTIO 128 \citep{rho_oph_spt_star}. Notably, this star exhibits a $\g$ of 18.52\,mag, is relatively close, situated at a distance of 136.3 $\pm$ 4.4 pc, and is young, with an estimated age of approximately 11 Myr \citep{upper_sco_age} and thus makes for a sensible choice when restricting the maximum brightness of UCDs. Drawing on the spectrophotometric correlations delineated by \cite{2019AJ....157..231K}, we infer that an L6 spectral type object in the G-$\grp$ domain should register a value of 1.77\,mag ($\g$-$\grp$ $\sim$ 1.47\,mag for an M7). Given that a substantial fraction of the objects discernible to both $\gaia$ and the UCDC likely belong to a preceding spectral category, this range was deemed appropriate for our constraints. As shown in Fig. \ref{blended}, the objects that fall within this demarcated region clearly deviate from the broader catalogue trends, revealing themselves as potential outliers. These objects are outlined in Table \ref{blended_table}, along with the details of their associated companions.
Several objects highlighted in Table \ref{blended_table} have been the subject of previous studies and are outlined as follows:

(i) (\textbf{J2200-3038A + J2200-3038B}): This is a known M9 + L0 resolved binary with a separation of $\rho$ = 1.09\arcsec ($\approx$ 35AU), as detailed in \cite{DENIS_J220002.05-303832.9AB}. This system was observed using NASA's Infrared Telescope Facility (IRTF) SpeX instrument \citep{SpeX}, and both objects were resolved on the north-south axis, with the northern component being brighter in all MKO bands. \cite{2019MNRAS.485.4423S} investigated this system as it was an outlier residing 3$\sigma$ away from the MS in $\g$-$\grp$ CMD. It was concluded that $\grp$ for the brighter component was a combination of fluxes from both objects in the system. The high-resolution imaging of this system reported by \cite{DENIS_J220002.05-303832.9AB} unequivocally confirms it as a bound double system with no other nearby sources, making the conclusion by \cite{2019MNRAS.485.4423S} the most probable explanation.

(ii) (\textbf{J1550+1455 + J1550+14553}): 2MASS J15500845+1455180AB is a well-studied binary system comprising of an L3.5 and L4 dwarfs, first discussed by \cite{2Mass_J15500845+1455180AB}. SDSS \citep{1999RSPTA.357...93M} initially classified this object as a galaxy due to the extended PSF in the \textit{i} and \textit{z} bands, which aligns with a faint, marginally discernible companion towards the north. 

(iii) (\textbf{J0025+4759A + J0025+4759}): J0025+4759 is a well-known and extensively studied equal-mass L4+L4 binary system that was initially resolved and characterised by \cite{2006AJ....132..891R}. However, this has not been resolved by $\gaia$, implying that $\grp$ is derived from the flux contributions from both sources. J0025+4759 is also known to be a distant companion to the spectroscopic binary J0025+4759A first discovered by \cite{2006AJ....132..891R}.

(iv) (\textbf{J0219-3925A + J0219-3925B}): The M6 J0219-3925A is a member of the Tucana-Horologium association (30-40 Myr) with a low-gravity L4$\gamma$ companion. The two objects are separated by 4$\arcsec$  with masses of 113 $\pm$ 12 M$_{\rm{Jup}}$ and 13.9 $\pm$ 1.1 M$_{\rm{Jup}}$ respectively, as reported by \cite{2MASS_J02192210-3925225}. The error in $\grp$ for the companion is large ($\sigma$$_{\grp}$ = 0.121\,mag), which is consistent within 1$\sigma$ of what we expect an L4 to be in $\g$ - $\grp$.

(v) (\textbf{J0753-6338 + J0753-6338B}):
\cite{2007MNRAS.378.1328M} classified J0753-6338B as having its own close companion consisting of an M7/8 and L2/3 dwarfs, which is unresolvable to $\gaia$, explaining a large number of blended transits and overall redness.

(vi) (\textbf{J0903-0637 + J0903-06378}):
Initially observed as a single object in 2MASS (2MASS J09033514-0637336), it was classified as an M7 in \cite{2003AJ....126.2421C}. However, \cite{2018ApJS..234....1B} serendipitously discovered this is a tightly bound visual binary system. Both objects have been resolved in $\gaia$ and PS1; the $\textit{i}$ images show a flux difference between the objects of 0.10 $\pm$ 0.03\,mag, and the difference in $\g$-band is $\sim$\,0.11\,mag \citep{2018ApJS..234....1B}.

(vii) (\textbf{J0915+0422A + J0915+0422}):
Initially classified as an L5 dwarf, \citep{2007AJ....133..439C}, J0915+0422 was subsequently imaged using the Hubble Space Telescope, revealing a closely separated binary system. This system consists of two nearly equal luminosity components, both of which were initially classified as L4. A follow-up mid-resolution spectral analysis in the optical range was conducted reclassifying both objects as L6 \citep{2008AJ....136.1290R(5)}. Further investigations using the SpeX spectrograph and low-resolution spectra (R $\sim$ 75-120) \citep{2014AJ....147...34S} led to the reclassification of the system as an L7, however, these observations did not resolve both components.

(viii) (\textbf{J1239+5515A + J1239+5515}):
J1239+5515 is a closely separated binary system ($\rho = 0.21\arcsec$) with an L5 combined NIR spectral type as reported by \cite{2003AJ....125.3302G}, which is unresolved in $\gaia$.

(ix) (\textbf{J1735+2634 + J1735+2634B}):
J1735+2634 was initially identified as a binary system by \cite{2006MNRAS.368.1917L} with subsequent mass estimations for the individual components produced in \cite{2017ApJS..231...15D}. Given that $\gaia$ has not resolved the low-mass secondary, the observed excess in the $\grp$ band is likely a cumulative contribution from both the components of the system.\\

(x) \textbf{J2331-0405}: 
J2331-0405 (HD 221356D), first discovered and characterised as an L1 dwarf by  \cite{2012MNRAS.427.2457G} is a well-known companion to HD 221356, forming a quadruple system with the binary system 2MASSW J2331016-040618, first resolved by \cite{2002ApJ...567L..53C}. J2331-0405 does not indicate unresolved binarity from either the IPD or RUWE values or any blended transits. The 2MASS photometry flags indicate that the photometry is unreliable (UUB), and the VHS photometry from \cite{2012MNRAS.427.2457G} appears bluer than expected for an L1 with a J-K$\rm{_s}$ = 1.008 \citep[J-K$\rm{_s}$ $\sim$ 1.32 for an L1, ][]{2019AJ....157..231K}. This can be partly explained by the fact that HD 221356 is slightly metal-poor ([Fe/H] = -0.26).

The systems presented below have not been identified with close companions:

(i) \textbf{J0112-7031 + J0112-7031B}: This system, recognised as 2MASS J01122168-7031235 \citep{2020yCat....102026M} with an M7 spectral type, as detailed in \cite{2003AJ....126.2421C}. Considering the small separation between the two companions and the absence of other celestial bodies nearby, it is plausible that these objects mutually contribute to their blended appearance.

(ii) \textbf{J0657-40197 + J0657-4019}:
J0657-4019 was previously classified as an M7.5, and reported for its moderate H$\alpha$ variability \citep{2003AJ....126.2421C, 2010ApJ...708.1482L}. J0657-4019 is not acknowledged as a binary system in the current literature. However, $\gaia$ data reveals two closely situated sources, both of which are potentially subjected to blending effects due to their proximity to each other ($\rho$ = 0.80$\arcsec$). This suggests the need for high-resolution follow-up observations to adequately identify and characterise both components. Given the nearly identical $\g$-band magnitudes of these sources, this system may be an equal-mass binary system.

(iii) \textbf{J1250+04553 + J1250+0455}:
J1250+0455 is classified as an M8.5 (Cheng et al, in prep), and found as a  wide companion with J1250+04553. Photometric spectral classification is estimated as an L1 \citep{2015A&A...574A..78S}, indicating a slight discrepancy between its spectral and photometric classification. The spectral features of J1250+0455 give further indications that this is not a typical late-M type object as the FeH absorption features at $\sim$1.0~$\mu$m and $\sim$1.2~$\mu$m are weaker than expected, the H$_2$O features at $\sim$0.9~$\mu$m and $\sim$1.1~$\mu$m are stronger than for a typical late-M UCD. The separation between the companions cannot explain the unusual redness exhibited by J1250+0455, with no other nearby $\gaia$ source suggesting a hidden companion contributing to the additional excess flux. This assertion aligns with the high RUWE and significantly large IPD flags. 

(iv) \textbf{J1317+18491 + J1317+1849}:
J1317+18491 is classified as an M9V by  \cite{2011AJ....141...97W}. $\gaia$ reveals a secondary source, J1317+1849, nearby ($\rho$ = 0.70$\arcsec$), indicating a tight companion system. This interpretation is further supported by the significantly large IPD flags and RUWE for both sources. All the transits for both sources are blended, explaining the considerable redness observed for both objects. Given the similarity in the $\g$-band for both components, we infer that this system is a near equal-mass binary.

(v) \textbf{J1347-76101 + J1347-7610}:
Initially classified as L0.0 in the NIR \citep{2007MNRAS.374..445K}, J1347-7610 appears considerably redder in $\g$-$\grp$ despite the small percentage of blended transits. The J-H value from 2MASS aligns with the L0 classification
\citep[J-H $\approx$ 0.76,, ][]{2019AJ....157..231K}. However, the \textbf{ipd\_frac\_multi\_peak} and RUWE values suggest an unseen secondary source semi-resolved by $\gaia$.
 
(vi) \textbf{J1517-5851A + J1517-5851}:
J1517-5851 is a confirmed wide BD companion to J1517-5851A \citep{2015MNRAS.454.4476S}. J1517-5851 has a J-K$\rm{_s}$ value of 1.33\,mag which resides between the average values for L1 and L1.5 dwarfs \citep{2013AJ....145....2F}, and its corresponding J, H, and K$\rm{_s}$ values align with the mean values of these dwarf classes \citep{2012ApJS..201...19D}. These measurements rule out the possibility of an unresolved hidden companion of a similar spectral type. Although there are no clear indications of blending or binarity, it resides in a densely populated stellar region, potentially contributing to the unusual flux observed in $\grp$.

\begin{table*}
\caption{Systems with at least one unusually red object as detailed in Sec.\ref{blended section} and displayed in Fig. \ref{blended} with their respective companions. $\grp$ obs - 	Number of observations (CCD transits) that contributed to the integrated RP mean flux, $\grp$ blended - Number of RP blended transits }
\label{blended_table}
\begin{tabular}{@{}lllllllllll@{}}
\toprule
SN &  ID & $\rho$ & SpT & $\g$ & $\g$-$\grp$ & amplitude & multi  & RUWE & $\grp$  & $\grp$    \\ 
&  & (\arcsec) &  &(mag) & (mag) &  & peak  &   & obs & blended  \\\midrule
J0025+4759A & 392562179817077120 & \multirow{2}{*}{217.5} & F8 &7.41 & 0.51 & 0.02 & 22 & 18.41 & 54 & 30  \\
J0025+4759 & 392555995065552000 &  & L4 &20.84 & 2.20 & 0.18 & 29 & - & 0 & 0  \\
\hline
J0112-7031 & 4690935350520872064 & \multirow{2}{*}{1.36} & M7 &18.27 & 1.94 & 0.02 & 1 & 1.10 & 25 & 21  \\
J0112-7031B & 4690935350519379712 &  & - & 18.29 &  2.01 & 0.03 & 0 & 1.05 & 17 & 15  \\
\hline
J0219-3925A & 4963614887043956096 & \multirow{2}{*}{3.91} & M6 & 14.98 & 1.40 & 0.01 & 0 & 2.92 & 72 & 3  \\
J0219-3925B & 4963614887043331072 &  & L4 & 21.05 & 1.82 & 0.05 & 0 & - & 7 & 0  \\
\hline
J0657-40197 & 5563853506012603776 & \multirow{2}{*}{0.80} & M7.5 &17.82 & 2.24 & 0.03 & 20 & 1.19 & 26 & 26  \\
J0657-4019 & 5563853506009853568 &  & - & 17.89 & 2.31 & 0.12 & 7 & 1.33 & 8 & 8  \\
\hline
J0753-6338 & 5287961677550103424 & \multirow{2}{*}{7.24} & G5V & 7.81 & 0.51 & 0.03 & 0 & 1.09 & 61 & 0 \\
J0753-6338B & 5287961677547964928 &  & M7 & 17.82 & 1.94 & 0.05 & 2 & 1.13 & 43 & 35  \\
\hline
J0903-0637 & 5756675238129804032 & \multirow{2}{*}{1.14} & M7 &18.30 & 1.99 & 0.11 & 20 & 2.63 & 39 & 39  \\
J0903-06378 & 5756675238130543104 &  & - &18.39 & 1.84 & 0.04 & 3 & 1.09 & 3 & 3 \\
\hline
J0915+0422A & 579379032258351488 & \multirow{2}{*}{2.11} & L6 &20.43 & 2.26 & 0.07 & 3 & 1.05 & 17 & 17  \\
J0915+0422 & 579379032258066432 &  & L7 & 20.58 & 2.45 & 0.24 & 0 & - & 4 & 4  \\
\hline
J1239+5515A & 1571411233756646656 & \multirow{2}{*}{2038} & K3V & 7.99 & 0.66 & 0.01 & 0 & 0.97 & 55 & 0  \\
J1239+5515 & 1571292108543541504 &  & L5 & 20.43 & 2.12 & 0.43 & 14 & - & 48 & 1  \\
\hline
J1250+04553 & 3705763723623026304 & \multirow{2}{*}{10.5} & - & 13.56 & 1.11 & 0.04 & 0 & 1.06 & 65 & 1  \\
J1250+0455 & 3705763723623660416 &  & M8 & 20.46 & 2.19 & 0.15 & 8 & 1.51 & 50 & 1 \\
\hline
J1317+18491 & 3938910834965369984 & \multirow{2}{*}{0.70} & M9V & 17.27 & 2.22 & 0.26 & 39 & 1.73 & 21 & 21 \\
J1317+1849 & 3938910830669894016 &  & - & 17.28 & 2.23 & 0.23 & 40 & 2.51 & 17 & 17  \\
\hline
J1347-76101 & 5789842624864892288 & \multirow{2}{*}{17.1} & M1 & 10.80 & 0.97 & 0.01 & 0 & 0.98 & 45 & 0 \\
J1347-7610 & 5789842620561789568 &  & L0 & 18.76 & 1.89 & 0.04 & 24 & 1.97 & 35 & 3 \\
\hline
J1517-5851A & 5877059048308526720 & \multirow{2}{*}{217.6} & A3 & 4.06 & 0.11 & 0.03 & 0 & 1.91 & 52 & 0 \\
J1517-5851 & 5877057330286155648 &  & L1 & 19.47 & 1.81 & 0.03 & 1 & 1.30 & 31 & 31  \\
\hline
J1550+1455 & 1192782138303692800 & \multirow{2}{*}{0.88} & L3.5 & 20.18 & 2.00 & 0.01 & 0 & 1.05 & 54 & 54  \\
J1550+14553 & 1192782134013894144 &  & - &20.87 & - & 0.06 & 0 & 1.10 & 0 & -  \\
\hline
J1735+2634 & 4594186745414679680 & \multirow{2}{*}{0.04} & M7.5 & 15.87 & 1.80 & 0.19 & 59 & 8.73 & 50 & 1  \\
J1735+2634B & - &  & L0  & - & - & - & - & - & - & - \\
\hline
J2200-3038A & 6616442994033876480 & \multirow{2}{*}{0.99} & M9  & 18.46 & 1.99 & 0.03 & 12 & 1.36 & 47 & 47  \\
J2200-3038B & 6616442994033876352 &   & L0 & 19.08 & - & 0.03 & 0 & 1.19 & 0 & -  \\
\hline
J2203-5647 & 6412595290592307840 & - & K5V & 4.32 & 0.71 & 0.005 & 0 & 1.148 & 45 & 0 \\
J2204-5646B & 6412596012146801152 & 75.70 & T1 & 18.06 & 2.00 & 0.01 & 8 & 4.30 & 44 & 42 \\
J2204-5646C & - & - & T6 & - & - & - & - & - & - & -  \\
\hline
J2331-0405A & 2633449134434620288 & 12.46 & F7V & 6.36 & 0.44 & 0.06 & 0 & 0.96 & 20 & 0  \\
J2331-0406 & 2633426216489127296 & 442.2 & M8 & 17.20 & 1.57 & 0.03 & 2 & 1.15 & 21 & 21  \\
J2331-0406B & - & 442.2 & L3 & - & - & - & - & - & - & - \\
J2331-0405 & 2633449095780158720 & - & L1 & 18.53 & 1.84 & 0.04 & 0 & 1.20 & 7 & 0 \\ \bottomrule
\end{tabular}
\end{table*}

\subsection{Binarity from Long Time-Baseline Astrometry}
\label{NSS_section}
Although Radial Velocity (RV) measurements or transits can provide accurate mass measurements, directly imaged UCDs and planets face a challenge in determining their mass. Direct imaging is sensitive to large projected separations between substellar objects and their host objects that correspond to longer orbital periods and larger semi-major axes. This makes it challenging to compute the masses of substellar mass companions, which often require sub-milliarcsecond precision \citep{2023A&A...674A..10H}. Precise astrometric measurements or RVs over a significant portion of the orbit are necessary to measure the mass of the directly imaged UCDs, which can be limited by the long orbital periods involved \citep{2022MNRAS.509.4411D}.

The availability of precise absolute astrometry from \textit{Gaia}, combined with pre-existing \textit{Hipparcos} astrometry, has enabled comparisons of widely spread astrometric measurements. By combining the PMs from both the \textit{Hipparcos} and \textit{Gaia} epochs, an acceleration (nonlinear PM) in the plane of the sky can be obtained. When combined with RV measurements, projected separation, and position angles, the data necessary to determine the dynamical mass without requiring observational data for a significant portion of the orbit are obtained \citep{2019AJ....158..140B}. This approach successfully determined masses of BD companions with long orbital periods, such as Gl 229 B ($\sim$240 years) \citep{2020AJ....160..196B} as well as shorter periods ($\sim$30 years) of $\beta$ Pic B \citep{betaPicB}.

$\gaia$ DR3 introduced astrometric orbital solutions that assist in determining the mass of companions, extending to the planetary mass regime. The $\gaia$ astrometric Non-Single Star (NSS) analysis includes sources that fail the single-star model; thus, a double-star model was used to determine the complete orbital solution for one companion \citep{2023A&A...674A..10H}. DR3 provides 169,227 NSS orbital astrometric solutions\footnote{https://www.cosmos.esa.int/web/gaia/dr3} that are compatible with an acceleration solution, found in the DR3 table \textbf{gaiadr3.nss\_acceleration\_astro}.

The GUCDS catalogue was passed through the NSS 'binary' pipeline to find detections of signals caused by substellar companions \citep{2023A&A...674A..34G}. The GUCDS \cite[an older version was used for this study,][]{2017MNRAS.469..401S}  was manually input into the binary pipeline as it only considers stars with $\g$ $\le$ 19\,mag.  This resulted in the determination of the orbital solution for 13 GUCDS objects, including 9 new binary candidates (these objects are categorised as nss\_solution\_type - OrbitalTargetedSearch), of which 2 are found in the UCDC: J2200-3038A  and J0219-3925A. These potential binary candidates are prime candidates for subsequent observations to characterise further and confirm their binarity. Their association with brighter primary stars in wide companion systems furthers their significance and positions them as valuable benchmarks for stellar studies. 
 
We crossmatched the \textbf{gaiadr3.nss\_acceleration\_astro} with the UCDC to determine if the accelerations detected are due to the UCD companions known in the UCDC, yielding 3 matches: J0122+0331A, J1449-0117A, and J1606+2253A. Using Kepler's third law, assuming a companion UCD mass of 0.1$M_\odot$ and inclination (\textit{i}) = 0, we estimate the orbital period for the systems. The periods are far larger ($>$ 100 years) than the 3.5-year baseline of $\gaia$ DR3, given the Hipparcos-Gaia baseline is best suited for detecting orbital periods of a few decades, whilst $\gaia$ is most effective at short rapid orbits with periods of several years, thus we cannot confidently say the accelerations detected are from our wide companions or an unresolved companion. 

A table of masses is derived from the NSS solutions presented in the \textbf{gaiadr3.nss\_two\_body\_orbit} table. The masses are listed in \textbf{gaiadr3.binary\_masses} and represent the estimated masses of the primaries and unresolved companions. We find 14 objects when cross-matching with the UCDC, all of which were non-UCDs. Some unseen companions have substellar masses at lower mass confidence levels (\textbf{m2\_lower} in \textbf{gaiadr3.binary\_masses}); for example, BD+75 510 which gives a lower companion mass of $\sim$ 0.006 \(\textup{M}_\odot\) but with large uncertainties, further data or follow-up observations are required to characterise the inner companion accurately.

The $\gaia$ Nearby Accelerating Star Catalogue (GNASC) \citep{2023AJ....165..193W} provides 29,684 high-confidence (5$\sigma$) accelerating star candidates within 100 pc from a supervised machine-learning algorithm trained on The Hipparcos Gaia Catalogue of Acceleration \citep{2021ApJS..254...42B}, DR2, and EDR3. The GNASC is sensitive to changes in the proper motion between DR2 and EDR3 astrometry, identifying objects not found in \textbf{gaiadr3.nss\_acceleration\_astro}. Notably, the GNASC still contains 96\% of all objects found in \textbf{gaiadr3.nss\_acceleration\_astro}, highlighting its coverage and effectiveness. There are 46 common stars between the UCDC and GNASC, of which 12 are UCDs with 3 new binary candidates listed in Table \ref{accelerating stars}. The \(\chi^2\) in Table \ref{accelerating stars} constitutes the difference between a linear drift model and the reported proper motions between the Hipparcos catalogue and EDR3 \citep{2021ApJS..254...42B}. A  \(\chi^2\) = 28.75 corresponds to a "5-$\sigma$" (99.7$\%$) confidence of non-linear motion, which the GNASC adopts for its selection criteria.

\begin{table}
\caption{Unresolved UCD companion candidates found in the GNASC \citep{2023AJ....165..193W} and also feature in the UCDC.  }
\label{accelerating stars}
\begin{tabular}{@{}llllll@{}}
\toprule
SN & SpT & multi\_peak & amplitude & RUWE & \(\chi^2\) \\ \midrule
J1453+1543 & M7.5 & 0 & 0.036 & 1.23 & 33.61 \\
J2147-26441 & M7.5e & 0 & 0.148 & 2.40 & 56.15 \\
J2325+4608A & M8 & 70 & 0.126 & 5.557 & 257.4 \\ \bottomrule
\end{tabular}
\end{table}

\section{Ages and Masses of UCDs}
\label{mass_age_section}
UCDs are characterised by their optical spectra, which exhibit strong and broad potassium lines, and their near-infrared spectra, which display the absorption bands of water, methane, and ammonia. These features are sensitive to [Fe/H] and surface gravity, both of which serve as age proxies \citep{2017MNRAS.470.4885M}. Most UCDs with stellar masses stabilise at a specific spectral type on the Main Sequence after a few Myrs. However, substellar objects are devoid of a long-term energy source, continuously cooling and evolving through subsequent spectral types.

Historically, constraining the age and mass of UCDs has proven challenging due to the continuous evolution and cooling of Brown Dwarfs, resulting in a degeneracy between their observed parameters and their age, mass, and metallicity. The dynamical mass of an astronomical object can be measured if it is part of a companion system with a short orbital period. However, the census of UCD masses remains relatively limited \citep{2017ApJS..231...15D, 2021AJ....162..301B, 2022AJ....163..288C, 2022ApJS..262...21F} and age indicators are typically the dominant uncertainty when estimating UCD masses \citep{2023ApJ...959...63S}. Members of Young Moving Group and the application of gyrochronology on host stars, however, can provide more accurate ages for low-mass stars \citep{2023ApJ...945..119G, 2021AJ....161...97C}. To estimate the masses of the UCDs, we first ascertain their ages and bolometric luminosities, and then interpolate the masses using the evolutionary tracks provided by \cite{2015A&A...577A..42B} for stellar objects and \cite{2020A&A...637A..38P} for substellar ($<$ 0.075$M\odot$) objects. Our interpolation method consists of calculating the minimum Euclidean distance between the nearest mass tracks and UCD. A weight that is inversely proportional to the minimum distance is then assigned to each track. Finally, we calculate the estimated mass as the weighted average of the two mass tracks, using the aforementioned weights.

\subsection{Bolometric luminosities}
We assume all systems are coeval and thus we infer the ages of UCDs in our sample from their primary companion. Given that we have accurate distance estimates for most UCDs from $\gaia$ parallaxes, we can compute the bolometric luminosity (L$\rm{_{bol}}$). We opt to use the L$\rm{_{bol}}$ over T$_{\rm{eff}}$ to infer mass as systematic differences remain between observed spectra and synthetic spectra persist due to complexities in the atmosphere, metallicity and age thus fitting observed spectra remain uncertain. L$\rm{_{bol}}$ integrates over all wavelengths and is less sensitive to model errors than flux corrected in a single bandpass.

We utilise the bolometric corrections (BCs) of \cite{2023AAS...24120311S} and in particular the $BC_{K_{s \text{FLD}}}$ relation for UCDs lacking evidence of youth or not known to be in Young Moving Groups (YMGs) and the $BC_{K_{\rm{s} \text{YNG}}}$
relation for those UCDs in YMGs. \cite{2023AAS...24120311S} highlights a concern using the J-band (both MKO and 2MASS) for BCs for young objects due to a large discrepancy between observed BCs and their best-fit relation for spectral ranges L5-L7. This discrepancy is not present when using the K$\rm{_s}$-band and thus we opted to use M$_{K_s}$ for our corrections. 

Since the bolometric magnitude, \(M_{\text{bol}}\), is:
\begin{equation}
M_{\text{bol}} = M_{K_s} + BC,
\end{equation}
it follows that the uncertainty on \(M_{\text{bol}}\), denoted as \(\sigma_{M_{\text{bol}}}\), is:
\begin{equation}
\sigma_{M_{\text{bol}}} = \sqrt{\sigma_{M_{Ks}}^2 + \sigma_{BC}^2},
\end{equation}
where \(\sigma_{M_{Ks}}\) and \(\sigma_{BC}\) are the uncertainties in the absolute magnitude in the K$\rm{_s}$ band and the BC, respectively.
 \(\sigma_{BC}\) is the rms around the fit as given by \cite{2023AAS...24120311S}.

Given the bolometric magnitude, \(M_{\text{bol}}\), and its uncertainty, \(\sigma_{M_{\text{bol}}}\), the bolometric luminosity, \(\log(L/L_{\odot})\), and its corresponding uncertainty, \(\sigma_{\log(L/L_{\odot})}\), are calculated as:
\begin{equation}
\log(L/L_{\odot}) = -0.4 \times (M_{\text{bol}} - M_{\text{bol},\odot}),
\end{equation}
\begin{equation}
\sigma_{\log(L/L_{\odot})} = 0.4 \times \sigma_{M_{\text{bol}}},
\end{equation}
where \(M_{\text{bol},\odot}\) is the solar bolometric magnitude (\(M_{\text{bol},\odot}\) = 4.74). Thus, the error in the luminosity ratio is directly proportional to the error in the bolometric magnitude, scaled by the factor of 0.4.

\begin{figure*}
\includegraphics[clip,width=2\columnwidth]{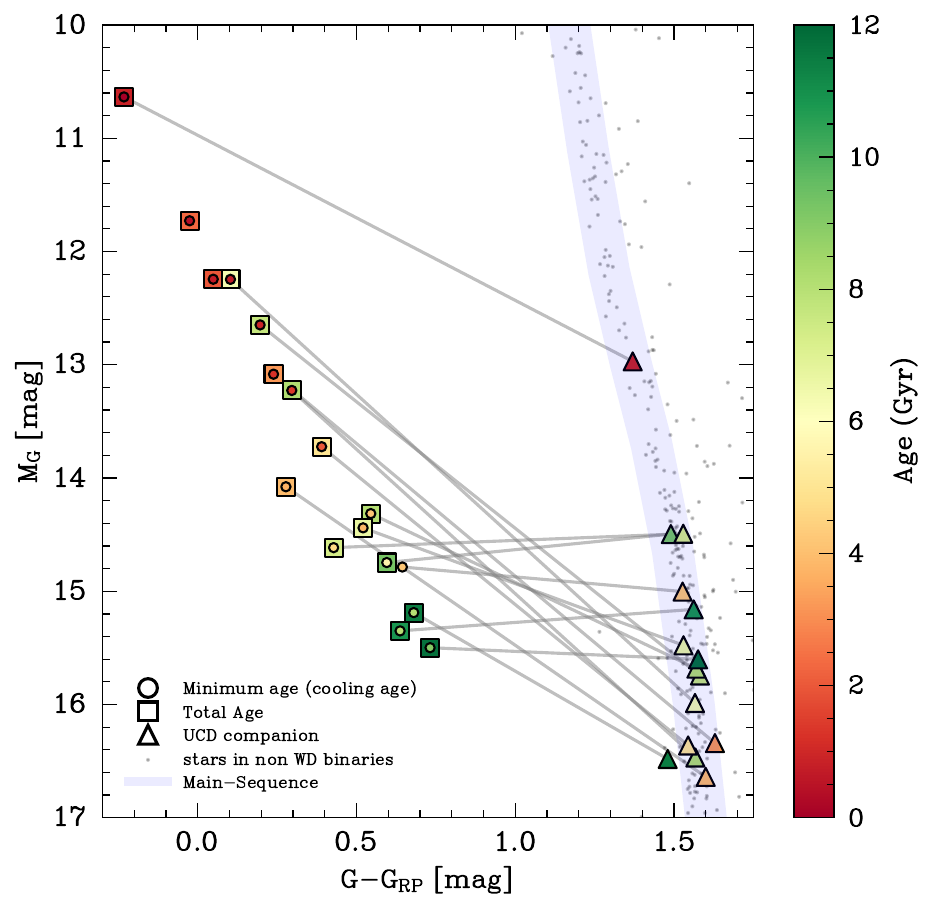}
\caption{Systems in the UCDC that have a WD and UCD component with minimum age and total age represented, where available. A summary of the data is presented in Table \ref{WD_age}. Objects in common systems are joined by grey lines. Circular markers indicate the minimum age (cooling age) and squared markers specify the total age. Triangular markers represent the UCD companion. The UCD age is represented by the total age of the WD companion unless only the cooling age can be determined, in which case, the UCD marker represents that age. J0807-6618A is represented without its companion as it is not seen by $\gaia$. }
\label{fig:WD_CMD}  
\end{figure*}

\subsection{Systems with a White Dwarf companion}
\label{WD section}
Accurately constraining the ages of cool and faint UCDs is challenging; however, one method to constrain UCD ages is to exploit resolved White Dwarf-UCD systems. By utilising photometric data, distance, and chemical composition of the WD, we can interpolate from WD cooling models to determine the age of the WD and, by extension, the UCD companion under the assumption of coevality \citep{WD_cooling, WD_cool2, 2019ApJ...870....9F}. In cases where the progenitor star has a large mass, it will have a short lifetime and the age of the WD (and UCD) will be approximately the same as the WD cooling age. This age can be calculated accurately from existing cooling models and the Initial-Final Mass Relations (IFMR) \citep{2008MNRAS.388..838D}.
Adopting the criteria used in \cite{2018MNRAS.480.4884E} we define WDs as $M_G$ $>$ 3.25($\gbp$-$\grp$) + 9.63, and limit $\g$ - $\grp$ $<$ 0.7\,mag, assuming negligible extinction within 100\,pc; hence, no correction is made. We identified 17 WD-UCD systems, consisting of 14 doubles, 2 triples, and 1 higher-order system. In cases of triple or higher multiplicity, we considered only the WD and UCD.

\begin{table*}
\centering
\caption{Derived parameters for WDs and their UCD companions in the UCDC. In cases where a WD's spectral type is unknown in the literature, both DA and DB scenarios were tested. Instances in which '-’ appears indicate no solution could be found. T$_{\rm{eff}}$ - effective temperature of the WD, log(g) - surface gravity of the WD, 
Cooling Age - The cooling age of the WD, Total Age -  Age of the progenitor + cooling age, where the age of the progenitor is determined using an Initial-Final Mass Relations from  \citep{2018ApJ...866...21C}}
\label{WD_age}
\renewcommand{\arraystretch}{1.2} 
\begin{tabular}{cccccccccc}
\toprule
WD SN & WD SpT & T$_{\rm{eff}}$ & log(g) & WD Mass & Cooling Age & Total Age & UCD SN & UCD SpT & UCD Mass \\ 
 & & (K) & (dex) & (M$_\odot$) & (Gyr) & (Gyr) & &  & (M$_{\rm Jup}$) \\ \midrule

J1424+0917A & DA4.1$^1$ & 12390 $\pm$ 194 & 8.05 $\pm$ 0.02 & $\asymerr{0.64}{+0.01}{-0.01}$ & $\asymerr{0.39}{+0.02}{-0.02}$ & $\asymerr{2.32}{+2.5}{-0.69}$ & J1424+0917 & L4 &  $\asymerr{74.9}{+2.5}{-2.6}$\\[1ex]

J1455+3725 & DA$^2$ & 8042 $\pm$ 169 & 8.09 $\pm$ 0.06 & $\asymerr{0.64}{+0.03}{-0.03}$ & $\asymerr{1.23}{+0.13}{-0.11}$ & $\asymerr{3.22}{+3.7}{-0.87}$ &  J1454+3718 & L0 & $\asymerr{81.6}{+1.5}{-2.0}$\\ [1ex] 

\multirow{2}{*}{J1455-2757} & DA & 7192 $\pm$ 331 & 8.48 $\pm$ 0.11 & $\asymerr{0.89}{+0.07}{-0.07}$ & $\asymerr{3.70}{+0.62}{-0.71}$ & $\asymerr{3.93}{+0.57}{-0.62}$ & \multirow{2}{*}{J1453-2744} & \multirow{2}{*}{L0.0}  &  \multirow{2}{*}{$\asymerr{85.6}{+3.1}{-3.1}$} \\ 
& DB & 7057 $\pm$ 330 & 8.43 $\pm$ 0.12 & $\asymerr{0.82}{+0.09}{-0.11}$ & $\asymerr{3.27}{+0.63}{-0.86}$ & $\asymerr{3.68}{+0.63}{-0.53}$ & & \\[1ex] 

J0030-3739A & DA$^3$ & 7272 $\pm$ 100 & 7.95 $\pm$ 0.04 & $\asymerr{0.57}{+0.02}{-0.02}$ & $\asymerr{1.36}{+0.08}{-0.08}$ & $\asymerr{8.12}{+4.4}{-3.5}$ &  J0030-3739 & M9 & $\asymerr{86.2}{+2.2}{-3.1}$ \\ [1ex] 

J1245+1204A & DA$^4$ & 6612 $\pm$ 175 & 8.04 $\pm$ 0.09 & $\asymerr{0.62}{+0.05}{-0.04}$ & $\asymerr{1.97}{+0.32}{-0.22}$ & $\asymerr{5.01}{+5.1}{-1.6}$ &  J1245+1204 & L1  & $\asymerr{88.3}{+5.6}{-5.8}$ \\ [1ex] 

\multirow{2}{*}{J1600-2456A} & DA & 4359 $\pm$ 190 & 7.99 $\pm$ 0.17 & $\asymerr{0.61}{+0.05}{-0.04}$ & $\asymerr{8.99}{+0.89}{-1.00}$ & $\asymerr{11.72}{+1.97}{-1.13}$ & \multirow{2}{*}{J1600-2456} & \multirow{2}{*}{M7V}  & \multirow{2}{*}{$\asymerr{88.4}{+2.1}{-4.1}$} \\ 
& DB & 4412 $\pm$ 142 & 7.98 $\pm$ 0.17 & $\asymerr{0.61}{+0.05}{-0.04}$ & $\asymerr{7.42}{+0.49}{-0.50}$ & $\asymerr{10.57}{+2.74}{-1.51}$ & & \\[1ex] 

\multirow{2}{*}{J0154+4819} & DA & 9762 $\pm$ 118 & 7.97 $\pm$ 0.03 & $\asymerr{0.59}{+0.02}{-0.02}$ & $\asymerr{0.66}{+0.04}{-0.03}$ & $\asymerr{6.36}{+4.7}{-3.1}$ &  \multirow{2}{*}{J0154+4819B} & \multirow{2}{*}{M9}  & \multirow{2}{*}{$\asymerr{90.3}{+2.5}{-1.9}$}\\ 
 & DB & 9571 $\pm$ 121 & 7.87 $\pm$ 0.04 & $\asymerr{0.54}{+0.04}{-0.02}$ & $\asymerr{0.63}{+0.03}{-0.03}$ & $\asymerr{9.30}{+3.4}{-3.7}$ \\[1ex] 

J2354-3316B & DA$^1$ & 8560 $\pm$ 75 & 7.96 $\pm$ 0.02 & $\asymerr{0.58}{+0.01}{-0.01}$ & $\asymerr{0.90}{+0.03}{-0.03}$ & $\asymerr{7.68}{+4.3}{-3.4}$ &  J2354-3316 & M9 & $\asymerr{90.6}{+1.6}{-1.6}$ \\ [1ex] 

\multirow{2}{*}{J1257+3347} & DA & 4416 $\pm$ 691 & 8.06 $\pm$ 0.67 & $\asymerr{0.63}{+0.08}{-0.05}$ & $\asymerr{7.97}{+2.3}{-2.8}$ & $\asymerr{10.85}{+2.34}{-2.72}$ & \multirow{2}{*}{J1258+3336} & \multirow{2}{*}{M8V}  &  \multirow{2}{*}{$\asymerr{90.7}{+2.6}{-4.1}$} \\ 
& DB & 4430 $\pm$ 535 & 8.02 $\pm$ 0.66 & $\asymerr{0.62}{+0.08}{-0.05}$ & $\asymerr{7.36}{+1.1}{-1.4}$ & $\asymerr{9.87}{+2.9}{-1.7}$  \\[1ex] 

\multirow{2}{*}{J1330+1403} & DA & 5658 $\pm$ 220 & 8.10 $\pm$ 0.16 & $\asymerr{0.63}{+0.06}{-0.04}$ & $\asymerr{3.82}{+1.4}{-0.93}$ & $\asymerr{6.61}{+3.7}{-1.5}$ & \multirow{2}{*}{J1330+1353} & \multirow{2}{*}{M8} & \multirow{2}{*}{$\asymerr{91.6}{+2.6}{-2.8}$} \\ 
& DB & 5531 $\pm$ 215 & 8.02 $\pm$ 0.17 & $\asymerr{0.60}{+0.05}{-0.04}$ & $\asymerr{4.55}{+0.90}{-0.93}$ & $\asymerr{8.17}{+4.0}{-1.90}$ & & \\[1ex] 

\multirow{2}{*}{J1401-0223} & DA & 5481 $\pm$ 248 & 7.91 $\pm$ 0.19 & $\asymerr{0.61}{+0.05}{-0.04}$ & $\asymerr{4.14}{+1.5}{-1.1}$ & $\asymerr{8.05}{+3.9}{-2.1}$ & \multirow{2}{*}{J1401-0221} & \multirow{2}{*}{M8}  &  \multirow{2}{*}{$\asymerr{93.3}{+6.8}{-3.7}$} \\ 
& DB & 5386 $\pm$ 228 & 7.84 $\pm$ 0.18 & $\asymerr{0.47}{+0.11}{-0.08}$ & $\asymerr{3.13}{+1.5}{-0.81}$ & - & & \\[1ex] 

J1208+0845A & DA$^{5}$ & 4930 $\pm$ 77 & 7.88 $\pm$ 0.07 & $\asymerr{0.51}{+0.03}{-0.03}$ & $\asymerr{4.80}{+0.80}{-0.74}$ & $\asymerr{9.32}{+2.9}{-1.8}$ &  J1208+0845 & M9 & $\asymerr{96.4}{+2.6}{-3.5}$ \\ [1ex] 

\multirow{2}{*}{J1605-20011} & DA & 5125 $\pm$ 180 & 7.99 $\pm$ 0.15 & $\asymerr{0.60}{+0.05}{-0.04}$ & $\asymerr{5.57}{+1.3}{-1.3}$ & $\asymerr{9.27}{+3.3}{-1.9}$ & \multirow{2}{*}{J1605-2001} & \multirow{2}{*}{M7V} & \multirow{2}{*}{$\asymerr{100}{+2.8}{-4.0}$} \\ 
& DB & 5053 $\pm$ 157 & 7.94 $\pm$ 0.13 & $\asymerr{0.59}{+0.04}{-0.03}$ & $\asymerr{5.85}{+0.61}{-0.65}$ & $\asymerr{10.1}{+3.1}{-2.2}$ & & \\[1ex] 

\multirow{2}{*}{J1153+2854} & DA & 5647 $\pm$ 382 & 8.22 $\pm$ 0.26 & $\asymerr{0.64}{+0.10}{-0.05}$ & $\asymerr{4.62}{+2.0}{-1.5}$ & $\asymerr{7.27}{+3.1}{-1.9}$ & \multirow{2}{*}{J1153+2901} & \multirow{2}{*}{M8V}  & \multirow{2}{*}{$\asymerr{101}{+4.9}{-5.2}$}\\ 
& DB & 5512 $\pm$ 376 & 8.13 $\pm$ 0.27 & $\asymerr{0.62}{+0.07}{-0.04}$ & $\asymerr{4.78}{+1.3}{-1.3}$ & $\asymerr{7.75}{+3.4}{-1.8}$ & & \\[1ex] 

J0428+1658 & DA.2.0$^1$ & 24200 $\pm$ 486 & 8.12 $\pm$ 0.03 & $\asymerr{0.70}{+0.02}{-0.02}$ & $\asymerr{0.035}{+0.06}{-0.05}$ & $\asymerr{0.85}{+0.49}{-0.35}$ & J0426+1703 & M7V & $\asymerr{133}{+6.2}{-5.3}$\\ [1ex]

\multirow{2}{*}{J1256-62023} & DA & 4562 $\pm$ 182 & 7.91 $\pm$ 0.17 & $\asymerr{0.60}{+0.05}{-0.04}$ & $\asymerr{8.05}{+1.0}{-1.1}$ & $\asymerr{11.44}{+2.22}{-1.41}$ & \multirow{2}{*}{J1256-6202} & \multirow{2}{*}{sdL3$^{6}$} & \multirow{2}{*}{-}  \\ 
& DB & 4588 $\pm$ 145 & 7.90 $\pm$ 0.16 & $\asymerr{0.51}{+0.09}{-0.08}$ & $\asymerr{5.83}{+1.3}{-1.4}$ & - & & \\[1ex] 

J0807-6618A & DQ$^{7}$ & 10250$^{\rm{a}}$ & 8.06$^{\rm{a}}$ & 0.62$^{\rm{a}}$ & $\asymerr{0.67}{+0.04}{-0.04}$$^{\rm{a}}$ & $\asymerr{1.5}{+0.5}{-0.3}$$^{\rm{b}}$ &  J0807-6618 & Y1$^{8}$ &  $\sim$ 7$^{\rm{b}}$\\ [1ex] 
\bottomrule
\end{tabular}
\smallskip
\footnotesize{\\Spectral Type References : \\
$^1$ - \cite{2011ApJ...743..138G(1)} ,  $^2$ - \cite{2006ApJS..167...40E(6)}, $^3$ - \cite{2008MNRAS.388..838D}, $^4$ - \cite{2013ApJS..204....5K(9)},  \\ $^{5}$ - \cite{2010ApJS..190...77K(13)}, $^{6}$ - \cite{2019MNRAS.486.1840Z},  $^{7}$ - \cite{2007AJ....134..252S}, $^{8}$ - \cite{2015ApJ...799...37L}}
\smallskip

\footnotesize{a - \cite{2009AJ....137.4547S}, b - \cite{2012ApJ...744..135L}}

\end{table*}

\begin{figure*}
  \centering
  \begin{subfigure}{\textwidth}
    \centering
    \includegraphics[width=0.5\linewidth]{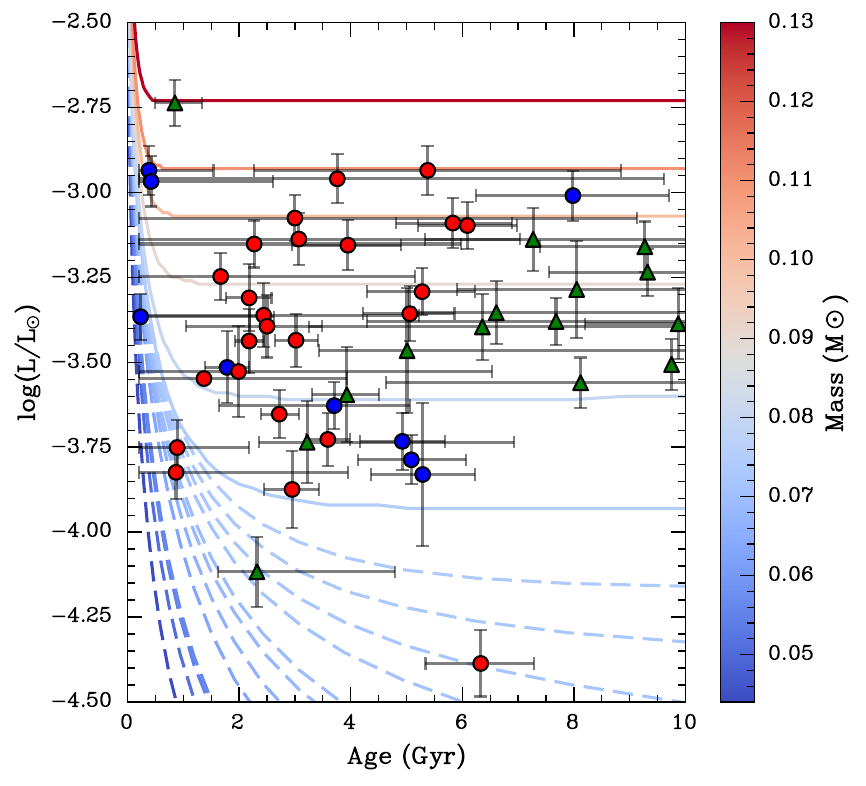}
    \caption{}
    \label{fig:7a}
  \end{subfigure}
  
  \begin{subfigure}{0.5\textwidth}
    \centering
    \includegraphics[width=\linewidth]{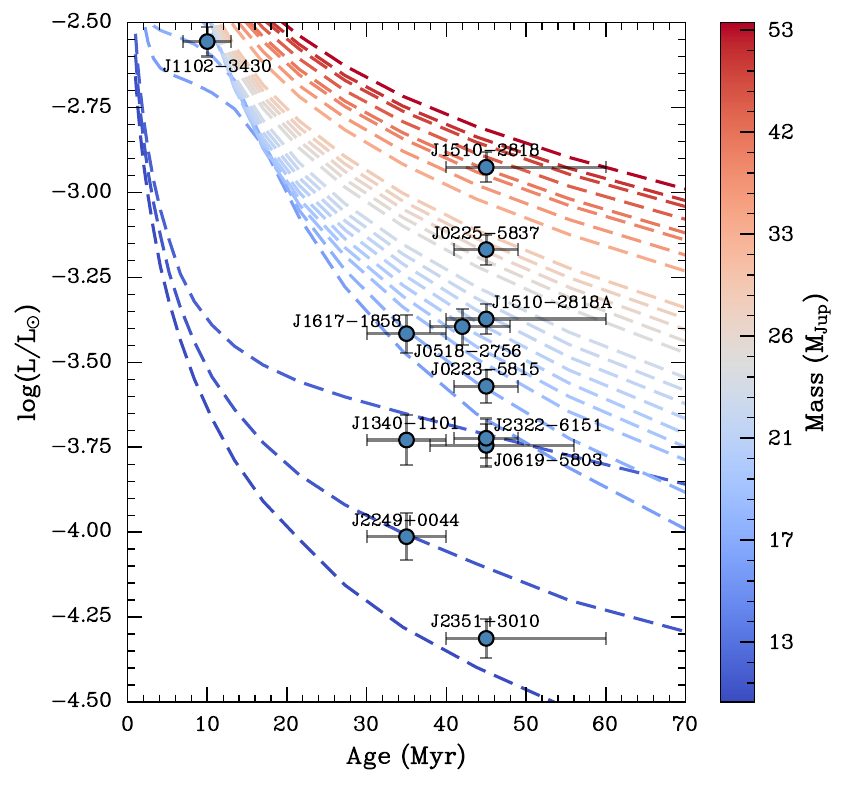}
    \caption{}
    \label{fig:7b}
  \end{subfigure}%
  \begin{subfigure}{0.5\textwidth}
    \centering
    \includegraphics[width=\linewidth]{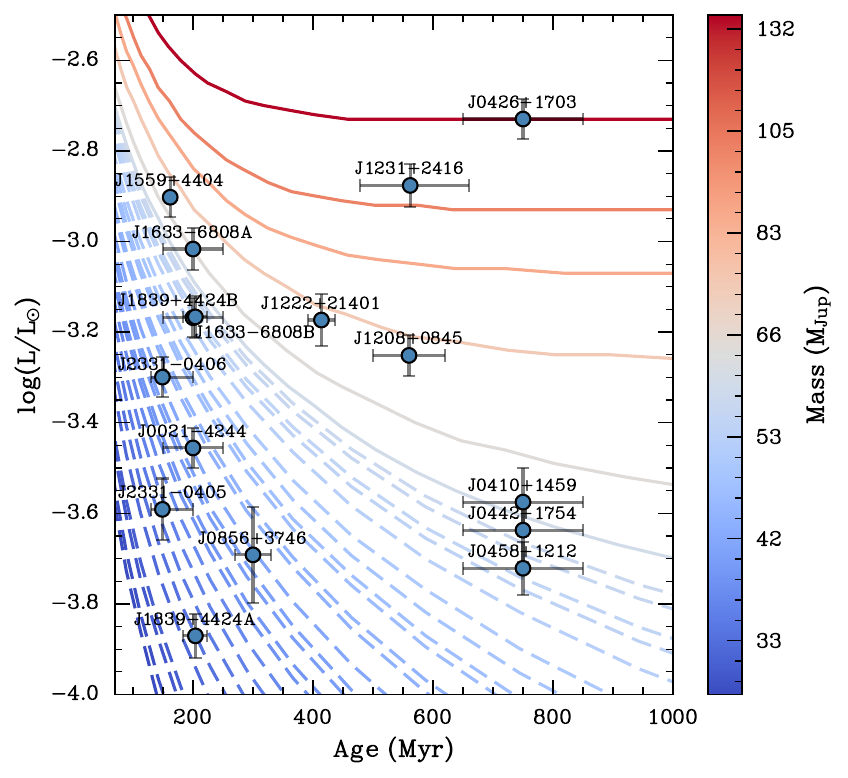}
    \caption{}
    \label{fig:7c}
  \end{subfigure}
  
  \caption{Evolutionary tracks for given masses and UCD positions in relation to $\log(L/L_{\odot})$ as a function of age. The dashed lines represent substellar mass tracks ($<$ 0.075$M\odot$) from \citep{2020A&A...637A..38P}, while solid lines depict stellar mass tracks  from \citep{2015A&A...577A..42B}. [Top (6a)]: UCDs with ages inferred from $\gaia$ FLAME where blue circular markers indicate GSP-Spec ages and red markers denote GSP-Phot ages. The green triangular markers denote UCDs with ages derived from a WD companion. [Bottom Left (6b)]: UCDs found in young moving groups from BANYAN $\Sigma$ with ages $<$ 70Myr. [Bottom Right (6c)]: UCDs found in young moving groups from BANYAN $\Sigma$ with ages $>$ 70Myr.  The arrangement of Figures 6b and 6c is chosen for readability purposes. \textbf{The objects in  6b and 6c are not included in 6a.}}
  \label{fig:7}
\end{figure*}

\subsubsection{UCD Ages from White Dwarf Companions}
\label{WD_ages_section}

The $\gaia$ mission has significantly improved our ability to accurately measure the photometry of WDs, highlighting the need to develop methods to easily transform WD photometry into physical parameters. To determine the cooling ages, we utilise the publicly available source code, \texttt{WDWARFDATE} \citep{2022AJ....164...62K}. \texttt{WDWARFDATE} uses a Basyiean framework to derive the age and masses of the WD and its progenitor based on its T$_{\rm{eff}}$ and  $\log\,g$, we use the values from \cite{2021yCat..75083877G} for these parameters. The cooling ages and mass of the WD are computed from the cooling models of the Montreal White Dwarf Group \citep{montreal_cooling}. 
The IFMR of \cite{2018ApJ...866...21C} is used to calculate the progenitor mass, and the lifetime of the progenitor (MS age) is determined from the Modules for Experiments in Stellar Astrophysics Isochrones and Stellar Evolution Tracks  \citep[MIST, ][]{2016ApJS..222....8D, 2016ApJ...823..102C}, 
thus giving the total age of the WD as the cooling age + MS age. The largest source of error on the total age is due to the IFMR as the large uncertainty from the initial mass correlates with the uncertainty in the MS age, comparatively, the $\gaia$ parallax and photometric uncertainties are negligible compared to the IFMR error.
We assumed solar metallicity and stellar rotation of \textit{v}/\textit{v}$_{\textit{crit}}$ = 0 for each fit. Table \ref{WD_age} displays the derived physical parameters of the UCDC WDs,  considering only pure hydrogen and helium (DA/DB) atmospheres with H-thick envelopes \( (\frac{M_{\text{H}}}{M_{*}} = 10^{-4} \))
and H-thin envelopes, respectively (\( \frac{M_{\text{H}}}{M_{*}} = 10^{-10} \)) as the Montreal cooling models are limited to these two scenarios. The Cummings-PARSEC \citep{2018ApJ...866...21C} IFMR from the PARSEC-isochrones \citep{2012MNRAS.427..127B} is adopted for J1256-62023. We adopt the IFMR of \citep{2009ApJ...693..355W} for the solutions of J1208+0845A. J0807-6618A has data from literature only.

Fig. \ref{fig:WD_CMD} displays the ages of the WDs and their UCD companions. The position of the UCD companion can confirm the age of a WD, as the eldest WDs are typically companions of the faintest and coolest UCDs in the sample. An example is J1256-62023 + J1256-6202, which is one of the oldest systems shown in Fig. \ref{fig:WD_CMD}. J1256-6202 displays a significant deviation from the MS, which indicates a sub-dwarf and metal-poor object. A follow-up study by \cite{2019MNRAS.486.1840Z} on J1256-6202  assigns a sdL3 spectral type,  firm halo membership, and [Fe/H] = -0.9.

\subsection{Using FLAME ages to estimate UCD masses}
\label{flame}
The Final Luminosity Age Mass Estimator (FLAME) provides a large, homogeneous, multi-parametric sample constraining stellar mass and evolutionary parameters for each $\gaia$ source ($\g$ $<$ 18.25\,mag)\footnote{\href{https://gea.esac.esa.int/archive/documentation/GDR3/Data_analysis/chap_cu8par/sec_cu8par_apsis/ssec_cu8par_apsis_flame.html}{https://gea.esac.esa.int/archive/documentation/GDR3/FLAME}}. FLAME processes the output spectroscopic parameters from the General Stellar Parametriser (GSP) from BP/RP and $\grvs$ spectra (GSP-Phot and GSP-Spec) with astrometry and photometry contributions to the output-derived evolutionary parameters (R, M, L, age) with upper and lower bound confidence levels. GSP-Phot aims to characterise all single stars ($\g$ $<$ 19\,mag) using time-averaged, low-resolution BP/RP spectra, parallax, and $\g$ \citep{2023A&A...674A..27A}. GSP-Spec is based solely on a spectroscopic approach that estimates stellar parameters from the combination of RVS spectra of single stars with no additional information from photometric or spectrophotometric BP/RP data \citep{2023A&A...674A..29R}. The FLAME results from GSP-Phot inputs are found in the \href{https://gea.esac.esa.int/archive/documentation/GDR3/Gaia_archive/chap_datamodel/sec_dm_astrophysical_parameter_tables/ssec_dm_astrophysical_parameters.html}{\rm{astrophysical\_parameters}} table while the outputs from GSP-Spec are in the  \href{https://gea.esac.esa.int/archive/documentation/GDR3/Gaia_archive/chap_datamodel/sec_dm_astrophysical_parameter_tables/ssec_dm_astrophysical_parameters_supp.html}{\rm{astrophysical\_parameters\_supp}} table. Where available we used the values from GSP-Spec for age estimations.

We cross-match the UCDC with the  \href{https://gea.esac.esa.int/archive/documentation/GDR3/Gaia_archive/chap_datamodel/sec_dm_astrophysical_parameter_tables/ssec_dm_astrophysical_parameters.html}{\rm{astrophysical\_parameters}} and \href{https://gea.esac.esa.int/archive/documentation/GDR3/Gaia_archive/chap_datamodel/sec_dm_astrophysical_parameter_tables/ssec_dm_astrophysical_parameters_supp.html}{\rm{astrophysical\_parameters\_supp}} tables, applying a cut where \textbf{age\_flame\_upper} $<$ 10 Gyr, as the evolutionary tracks from \cite{2015A&A...577A..42B} and \cite{2020A&A...637A..38P} used to interpolate the masses are limited to this age, resulting in 40 matches, 28 from GSP-Spec and 12 from GSP-Phot, tabulated in Table \ref{mass table}.\\

\subsection{\texorpdfstring{Using BANYAN $\Sigma$ ages to estimate UCD masses}{Using BANYAN Sigma ages to estimate UCD masses}}
\label{banyan}
Assuming coevality for objects formed in the same molecular cloud provides valuable age estimations for all members within the association, with a potential precision of a few Myr \citep{2015MNRAS.454..593B}.

BANYAN $\Sigma$ \citep{2020ApJ...903...96G} is a Bayesian algorithm created to identify members of nearby young stellar associations within 150\,pc with an age range $\sim$ 1-800\,Myr. The algorithm compares in six-dimensional galactic coordinates and space velocities (\textit{XYZUVW}) to the included associations to compute a membership probability of the associations and the field. It is worth noting that while  BANYAN $\Sigma$ is a robust tool for this task, other alternative tools are available, such as \texttt{Clusterix 2.0} \citep{2020MNRAS.492.5811B}.
$\gaia$ provides motions and positions for the UCDC and was thus queried using BANYAN $\Sigma$ to discover UCDs that belong to young stellar associations. To mitigate the false-positive rate (percentage of field stars erroneously classified as members), we focus on potential members with a membership probability threshold of $\geq$ 90$\%$.
BANYAN $\Sigma$ is capable of a 50$\%$ detection rate for true members considering only proper motion data. However, the inclusion of additional data, such as RVs and parallaxes, improves the confidence of detection at the same probability threshold. Specifically, it increases to 68$\%$ when RVs and proper motion are used, 82$\%$ when proper motion and parallax are considered and could peak at 90$\%$ when all three parameters are used.

Given the variances in size, distance, and membership completeness of Moving Groups, different probability thresholds are required to maintain equivalent true recovery rates across all groups. To resolve this association-specific dependence, BANYAN $\Sigma$ adjusts the Bayesian priors to unify the recovery rates for all Moving Groups. As a result, the probability threshold does not precisely reflect membership probability but rather serves as a functional value that enables uniform classification performance across all Moving Groups. As these associations are young, we expect no significant perturbation from other non-member stars to have occurred; thus, all members should share similar space velocities UVW with typical velocity dispersions $<$ 3\,kms$^{-1}$ \citep{2020ApJ...903...96G}. 
Objects for which the separation in UVW space is not closer than 3\,kms$^{-1}$ from the centre of the associated BANYAN model were excluded. This analysis results in 35 potential matches, once non-UCD companions are excluded, from which 28 UCD-derived mass estimations are obtained, as tabulated in Table \ref{banyan_table}. The 7 objects without mass estimates are due to a lack of an age estimate or $\log(L/L_{\odot})$. Evolutionary tracks in relation to the age and $\log(L/L_{\odot})$ of the UCDs discussed in Sec.\ref{mass_age_section} are presented in Fig. \ref{fig:7}. A similar analysis is conducted by \cite{2023A&A...674A..39G} using UCDs from \cite{2019MNRAS.485.4423S} with companions possessing FLAME ages, resulting in some overlap with our targets and exhibiting consistency between the masses derived from both studies.  \\

\subsection{Verification of ages}

\subsubsection{Verification of age using variability}
$\gaia$ DR3 provides time-series analysis for 10.5 million sources from the 34-month multi-epoch data summarised by \cite{2023A&A...674A..13E}. Variability in UCDs was explored in \cite{2023A&A...669A.139S} highlighting UCDs with asymmetric $\g$-band distributions that exhibit significant dips in brightness (assumed to be eclipsing binaries) and bright outliers (believed to be flares). Variability is often a useful indicator of youth, revealing information about their rotation and magnetic activity, with flaring UCDs in young Moving Groups that have been previously observed and characterised \citep{2017ApJ...845...33G}. 
The \textbf{gaiadr3.vari\_summary} table provides a summary of all the variable objects from all 17 variability tables in $\gaia$ DR3, upon cross-matching the \textbf{gaiadr3.vari\_summary} table with the UCDs listed in Table \ref{banyan_table}, we find that none of these objects are present in the variability summary. This result suggests that these particular UCDs either exhibit no noticeable variability within the observational timescale or their variability falls below the detection threshold of $\gaia$'s DR3 dataset. \\

\subsubsection{Verification of age from kinematics}
The tangential velocities (V$_{\rm TAN}$) for all UCDs are included in Table \ref{mass table} and Table \ref{banyan_table}. Halo stars typically move faster than disk stars as younger disk members demonstrate less kinematic activity than their older counterparts because of fewer tidal perturbations from nearby celestial objects. A notably large tangential velocity (V$_{\rm TAN}$ $\ge$ 100kms$^{-1}$) serves as a useful proxy for age, indicating potential low metallicity Halo stars. V$_{\rm TAN}$ is especially useful in hard-to-measure objects such as BDs where obtaining the RV for full three-dimensional motion is challenging. However, the analysis does not show a clear correlation between the V$_{\rm TAN}$ and age. The absence of a discernible correlation between V$_{\rm TAN}$ and the ages of UCDs provides reasons for caution regarding the accuracy of the $\gaia$ ages, however, it provides a statically useful homogeneous all-sky sample of ages. It should be noted that the use of V$_{\rm TAN}$ as a proxy for age may be somewhat limited, given that it captures only two dimensions of motion, lacking the Radial Velocity for a full three-dimensional kinematic picture.

We analyse the Galactic components of our UCDs discussed in Sec. \ref{mass_age_section} using a Toomre diagram with the usual criterion of Galactic velocity components, \textit{U} is towards the Galactic centre, \textit{V} is in the direction of the plane rotation and \textit{W} is perpendicular to the Galactic plane, positive towards the North Galactic Pole and are expressed with respect to the Local Standard of Rest. The \textit{UVW} velocities and their respective errors are determined using the \texttt{SteParKin} code \citep{2001A&A...379..976M, 2020A&A...642A.115C} with error propagation, following the method discussed in \cite{1987AJ.....93..864J}.

\begin{figure}
\includegraphics[width=\columnwidth]{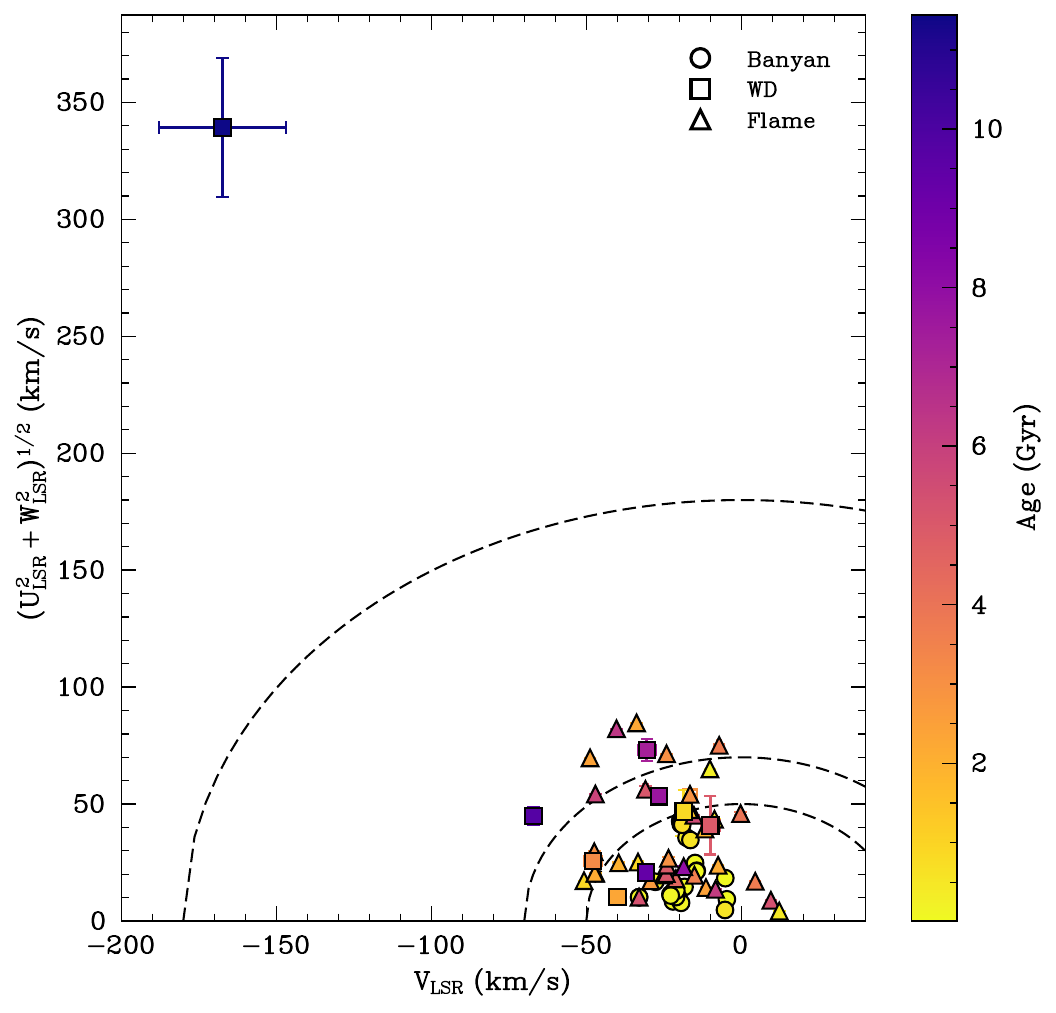}
\caption{Toomre Diagram for UCDs discussed in Sec.\ref{mass_age_section}. The dashed lines show the total space velocity, $v_{\text{tot}} \equiv \sqrt{U_{\text{LSR}}^2 + V_{\text{LSR}}^2 + W_{\text{LSR}}^2}$, at 50, 70 and 180 kms$^{-1}$ where sources $\le$ 50kms$^{-1}$ are typically thin disk, 70kms$^{-1}$ $\le$ $v_{\text{tot}}$ $\le$ 180kms$^{-1}$ are probable thick disk members  \citep{20d0f820213911dbbee902004c4f4f50, 2014A&A...562A..71B}.  }

\label{fig:Toomre Diagram}
\end{figure}

Fig. \ref{fig:Toomre Diagram} presents the kinematic distribution of the UCD sample, categorising each into thin-disc, thick-disc, or Halo populations provides a reliable indirect approach to deducing their age, given the distinct kinematic and age signatures of each population exhibits. The main drawback of this approach is the requirement of full 3-velocity components, \textit{UVW}, which is an extremely difficult task, given that RVs are required which is inherently difficult to obtain for UCDs. This issue is circumvented by using the space motions from the primaries, as RVs are widely available for brighter components. It is assumed that the UCDs have the same motions as their primary. The distributions shown in Fig. \ref{fig:Toomre Diagram} underscores this distinction, with younger UCDs from Moving Groups, as identified by BANYAN $\Sigma$, predominantly populating the thin-disc, whereas older counterparts from FLAME and WD ageing are more prevalent within the thick-disc or Halo realms. 

The use of space velocities offers a comprehensive view of the age distributions within our sample; however, individual cases require further attention. The thin disk is thought to have originated between 8 to 10 Gyr ago \citep{2016A&A...588A..35T, 2017ApJ...837..162K}, following a period of slowed major mergers and settling of gaseous components. This results in a large variance in age within the thin disk. For example, J1208+0845 has an estimated age exceeding 9\, Gyr, as ascertained from its WD companion and unambiguously situated in the thin disk from Figure \ref{fig:Toomre Diagram}; its companion is also verified as a thin disk member by \cite{2019MNRAS.485.5573T}. This is in contrast with younger YMG UCD members with sub 100\, Myr ages also existing in the thin-disk. The sole Halo member depicted in Figure \ref{fig:Toomre Diagram} is J1256-6202. As discussed in Section \ref{WD_ages_section}, this Halo membership is consistent with the age and sub-dwarf classification of J1256-6202.

\begin{table}
\caption{ UCDs with age and mass solutions, as seen in Fig. \ref{fig:7a}. Ages are from  $\gaia$ FLAME, ages with $\ast$ indicate the use of GSP-Phot, and other ages are from GSP-Spec. The mass is given by Jupiter masses. 3 objects were not included in this table as they had not been resolved in NIR sky surveys, lacking photometric data (M$_{\rm J}$) and thus no mass estimation. SpT(P) - Primary companion spectral type. }
\label{mass table}
\renewcommand{\arraystretch}{1.5}
\begin{tabular}{@{}cccccc@{}}
\toprule
SN & SpT & Mass & Age & V$_{\rm TAN}$  & SpT(P)\\
& & (M$_{\rm Jup}$) & (Gyr) & (kms$^{-1}$)    \\\midrule

J1423+0116 & T8p & $\asymerr{23.9}{+3.1}{-4.0}$ & $\asymerr{5.57}{+1.6}{-1.6}$  & 43.0 $\pm$ 0.19 & G1.5V  \\

J1646+5019 & M7 & $\asymerr{58.2}{+34}{-0.20}$ & $\asymerr{0.24}{+2.4}{-0.04}$$\ast$  & 61.9 $\pm$ 0.14 & K3  \\
 
J2144+1446 & T2.5 & $\asymerr{59.6}{+4.8}{-5.9}$ & $\asymerr{4.46}{+0.93}{-0.93}$$\ast$  & 21.7 $\pm$ 0.05 & G0V  \\

J1339+0104 & T5 & $\asymerr{68.4}{+1.9}{-1.9}$ & $\asymerr{4.63}{+0.73}{-0.73}$$\ast$  & 31.8 $\pm$ 0.78 & F7V  \\

J1110-2925   & L2   & $\asymerr{71.7}{+8.0}{-36}$ & $\asymerr{0.87}{+3.1}{-0.67}$ & 14.3 $\pm$ 0.25 & K5V  \\

J0122+0331 & L1 & $\asymerr{74.6}{+6.1}{-29}$ & $\asymerr{0.89}{+1.3}{-0.69}$  & 37.8 $\pm$ 0.72 & G5 \\

J1416+5006 & L5.5 & $\asymerr{75.5}{+0.90}{-0.90}$ & $\asymerr{6.33}{+0.96}{-0.99}$  & 75.3 $\pm$ 1.2 & G5 \\

J1450+2354C & L4   & $\asymerr{78.6}{+5.8}{-0.80}$ & $\asymerr{5.29}{+0.93}{-0.93}$$\ast$  & 12.8 $\pm$ 0.014 & F9IV   \\

J1112+3548B & L4.5 & $\asymerr{79.0}{+4.1}{-1.3}$ & $\asymerr{5.09}{+0.98}{-0.96}$$\ast$  & 31.4 $\pm$ 0.38 & G2 \\

J0223+5240 & L1.5 & $\asymerr{79.5}{+1.8}{-1.8}$ & $\asymerr{2.95}{+0.48}{-0.50}$ &  32.2 $\pm$ 1.62 & F5  \\

J1800+1505 & L1 & $\asymerr{81.2}{+1.1}{-1.8}$ & $\asymerr{3.59}{+0.40}{-0.39}$  & 24.3 $\pm$ 0.42 & G5IV \\

J1217+1427 & L1 & $\asymerr{81.6}{+0.80}{-1.4}$ & $\asymerr{4.93}{+0.76}{-0.76}$$\ast$  & 25.7 $\pm$ 3.4 & F8  \\

J1022+4114 & L0 & $\asymerr{82.3}{+2.5}{-1.9}$ & $\asymerr{2.72}{+0.35}{-0.33}$ & 33.6 $\pm$ 0.30 & F7V \\

J1632+3505 & L0.5 & $\asymerr{82.6}{+2.2}{-0.90}$ & $\asymerr{3.71}{+1.4}{-2.1}$$\ast$ & 17.7 $\pm$ 0.097 & K0 \\

J0900+3205 & L1.5 & $\asymerr{86.4}{+3.4}{-1.5}$ & $\asymerr{1.79}{+0.40}{-0.40}$$\ast$  & 33.1 $\pm$ 1.4 & F2  \\

J1030+0052 & M8 & $\asymerr{87.1}{+3.8}{-38}$ & $\asymerr{1.99}{+4.6}{-1.8}$  & 49.4 $\pm$ 2.2 &  - \\

J1450+4617 & L0 & $\asymerr{89.2}{+2.3}{-2.6}$ & $\asymerr{3.02}{+0.39}{-0.38}$ & 71.2 $\pm$ 1.15 & F8  \\

J1141+4116 & L0    & $\asymerr{89.4}{+2.1}{-2.3}$ & $\asymerr{2.18}{+0.27}{-0.26}$ & 21.6 $\pm$ 0.72 & F5 \\

J2244-00291 & M9 & $\asymerr{89.8}{+7.5}{-2.0}$ & $\asymerr{2.50}{+0.99}{-1.5}$ &   18.1 $\pm$ 0.61 & G5V \\

J0025+4759 & L4 & $\asymerr{91.4}{+1.7}{-2.4}$ & $\asymerr{5.06}{+0.80}{-0.84}$  & 57.2 $\pm$ 1.2 & F8  \\

J1518+5328 & M9 & $\asymerr{92}{+1.5}{-3.0}$ & $\asymerr{2.44}{+2.8}{-2.2}$ &  77.2 $\pm$ 1.71& -  \\

J1637+2443 & M9V & $\asymerr{92.7}{+3.5}{-3.4}$ & $\asymerr{2.18}{+0.40}{-0.41}$ &  54.7 $\pm$ 1.64 & F5 \\

J2331-0406 & M8 & $\asymerr{93.6}{+2.2}{-2.2}$ & $\asymerr{5.28}{+0.94}{-0.99}$ &  31.9 $\pm$ 0.069 & F7V \\

J1320+0957   & M7.5 & $\asymerr{95.9}{+3.9}{-30}$ & $\asymerr{1.67}{+3.5}{-1.5}$  &  49.6 $\pm$ 0.21 & K2 \\

J1606+2253 & M8 & $\asymerr{101}{+3.3}{-5.4}$ & $\asymerr{3.95}{+0.95}{-1.0}$ & 49.2 $\pm$ 0.52 & G0  \\

J1204+3437   & M7   & $\asymerr{101}{+2.4}{-26}$ & $\asymerr{2.27}{+3.7}{-2.1}$ & 85.5 $\pm$ 0.53 &  K4 \\

J1421+30551   & M8v   & $\asymerr{101}{+6.4}{-16}$ & $\asymerr{3.07}{+4.0}{-2.9}$ &  43.0 $\pm$ 0.41 & -\\

J2226-7503 & M8 & $\asymerr{103}{+3.9}{-3.3}$ & $\asymerr{6.09}{+0.89}{-0.88}$ &  6.7 $\pm$ 0.0013 & G0V  \\

J0255+2136   & M7.5 & $\asymerr{104}{+5.5}{-3.2}$ &$\asymerr{5.83}{+1.1}{-1.0}$& 63.6 $\pm$ 0.77 & F8 \\

J1210+1858   & M7   & $\asymerr{107}{+3.9}{-29}$ & $\asymerr{3.00}{+6.1}{-2.8}$ &  35.6 $\pm$ 0.11 & K3V  \\

J1346+2208 & M9V & $\asymerr{109}{+6.4}{-3.2}$ & $\asymerr{0.42}{+2}{-0.2}$$\ast$  & 13.2 $\pm$ 0.23 & - \\

J1606+1748 & M7V & $\asymerr{110}{+4.0}{-6.0}$ & $\asymerr{7.98}{+1.7}{-1.7}$$\ast$ &  20.5 $\pm$ 0.24 & - \\

J2307+0520 & M7 & $\asymerr{114}{+3.4}{-5.6}$ & $\asymerr{3.76}{+5.9}{-3.6}$ & 71.2 $\pm$ 0.78& - \\

J0835+1714 & M7V & $\asymerr{115}{+8.9}{-4.4}$ & $\asymerr{5.38}{+3.5}{-3.1}$ &  39.7 $\pm$ 0.47 & K0  \\ \bottomrule

\end{tabular}
\end{table}

\section{Discussion} 
\label{conclusion}
Using $\gaia$ DR3 and a combination of the GCNS and GUCDS, we identify $\nsys$ multiple systems with at least one spectroscopically determined UCD, derived from astrometric constraints at a range of separations($\sim$ few AU - 100,000\,AU). The catalogue includes diverse samples of objects spanning spectral types (M7-Y1), metallicities, and ages. We also provide a detailed description of the selection criteria and methodology used in creating the catalogue to ensure its transparency and reproducibility as discussed in Sec.\ref{cat creation} and present an example system in Table \ref{Table1}. We analyse potential unresolved UCDs in our sample in Sec.\ref{indentify binaries}, identified using the LUWE and colour excess from blended sources (additional flux from a nearby secondary source contributing to the UCDs measured flux). In Sec.\ref{mass_age_section} we use the coeval nature of binaries to estimate the ages of 19 UCDs from their White Dwarf companion using WD cooling and IFMR models. $\gaia$ DR3 FLAME module provides ages of 40 UCDs and BANYAN $\Sigma$ finds 34 UCDs in Young Moving Groups with age estimates. The masses for these UCDs are estimated using current low-mass evolutionary models \citep{2015A&A...577A..42B, 2020A&A...637A..38P}.

The UCDC's systems are benchmark UCDs. Inferring the age and composition obtained from the primaries provides direct tests of current substellar models to help accurately define the hydrogen-burning limit and untangle the current age-mass degeneracy of BDs. The catalogue can also serve as a reference for surveys and observational programs targeting UCDs, allowing for more efficient and effective use of observation time.

We have discussed some uses of the UCDC, however, as this study was primarily concerned with the creation of the UCDC below we note some other possible uses for this sample:
\begin{itemize}[noitemsep,topsep=1pt]
\item\textit{Search for unresolved binarity}: This study demonstrates that some of our systems could harbour substellar companions that make prime follow-up imaging targets. If indeed there are unresolved companions, inferences on age and composition can be made based on the wide primary companion, if bright enough. 
\item\textit{Calibration of UCD ages}: We establish age estimations for 86 UCDs which in turn can be used to calibrate poorly calibrated age indicators for UCDs, which is especially useful when constraining field UCD ages.
\item\textit{Constraining the UCD binary fraction}: UCD binary and multiplicity fractions have been explored in previous literature, see \cite{2019ApJ...883..205B, 2020MNRAS.499.5302D}. This sample of UCD companion systems within 100\,pc can be used to constrain the general multiplicity fraction. 
Gravitational kicks from nearby objects often perturb and break apart binary systems over time. Using V$_{\rm TAN}$ as an age proxy, the relationship between age and multiplicity can be studied \citep[e.g. see][and references therein]{2019AJ....157..216W}.
\end{itemize}

\begin{table}
\caption{Overview of UCDs in young stellar associations identified using BANYAN $\Sigma$. Masses are derived from Fig.\ref{fig:7b} and Fig.\ref{fig:7c}.}
\renewcommand{\arraystretch}{1.5}
\label{banyan_table}
\begin{tabular}{@{}llllll@{}}
\toprule

SN & SpT & Mass & Age & Group  & V$_{\rm TAN}$  \\
& & (M$_{\rm Jup}$) & (Myr) &  & (kms$^{-1}$)  \\\midrule

J2351+3010 & L5.5 & $\asymerr{11.9}{+0.40}{-0.20}$ & $\asymerr{45}{+15}{-5}$ & ARG $^2$  &  29.6 $\pm$ 0.48   \\

J2249+0044 & L3$\gamma$ & $\asymerr{12.5}{+0.20}{-0.20}$ & $\asymerr{35}{+5}{-5}$ & OCTN$^1$  & 15.1 $\pm$ 0.48    \\

J1340-1101 & L1 & $\asymerr{13.4}{+0.30}{-0.20}$ & $\asymerr{35}{+5}{-5}$ & OCTN  & 12.0 $\pm$ 0.42  \\

J0619-5803 & L1.0 & $\asymerr{13.5}{+0.20}{-0.20}$ &$\asymerr{45}{+11}{-7}$ & CAR $^3$  & 10.3 $\pm$ 0.26   \\

J2322-6151 & L2$\gamma$ & $\asymerr{13.6}{+0.30}{-0.20}$ & $\asymerr{45}{+4}{-4}$ & THA$^3$ &  22.9 $\pm$ 0.47   \\

J1617-1858 & s/sdM7 & $\asymerr{14.8}{+0.90}{-0.30}$ & $\asymerr{35}{+5}{-5}$ & OCTN  &  6.14 $\pm$ 0.21  \\

J0223-5815 & L0$\gamma$ & $\asymerr{22.1}{+2.0}{-2.1}$ & $\asymerr{45}{+4}{-4}$ & THA &  20.0 $\pm$ 0.19  \\

J1102-3430 & M8.5$\gamma$ & $\asymerr{25.3}{+10}{-4.4}$ & $\asymerr{10}{+3}{-3}$ & TWA$^3$ &  19.9 $\pm$ 0.08  \\

J0518-2756 & L1$\gamma$ & $\asymerr{26.2}{+1.1}{-1.8}$ & $\asymerr{42}{+6}{-4}$ & COL$^3$ & 8.86 $\pm$ 0.19  \\

J1510-2818A & M9 & $\asymerr{28.1}{+9.8}{-1.9}$ & $\asymerr{45}{+15}{-5}$ & ARG  &  21.5 $\pm$ 0.14  \\

J0225-5837 & M9 & $\asymerr{34.6}{+2.0}{+1.1}$ & $\asymerr{45}{+4}{-4}$ & THA & 19.9 $\pm$ 0.08  \\

J1839+4424A & L2 & $\asymerr{38.7}{+1.8}{+1.6}$ & $\asymerr{185}{+20}{-20}$ & {Theia 213}$^4$ & 30.1 $\pm$ 0.30\\

J2331-0405 & L1 & $\asymerr{42.1}{+5.9}{-3.0}$ & $\asymerr{149}{+51}{-19}$ & ABDMG  &  31.8 $\pm$ 0.15  \\

J1510-2818 & M9 & $\asymerr{46.1}{+3.1}{-2.1}$ & $\asymerr{45}{+15}{-5}$ & ARG &  22.2 $\pm$ 0.08  \\

J0856+3746 & M8 & $\asymerr{53}{+4.5}{-4.6}$ & $\asymerr{300}{+30}{-30}$ & Crius 198$^{4}$  & 32.0 $\pm$ 1.40  \\

J0021-4244 & M9.5 & $\asymerr{53.7}{+5.5}{-6.6}$ & $\asymerr{200}{+50}{-50}$ & CARN$^5$ &  32.2 $\pm$ 0.16  \\

J2331-0406 & M8 & $\asymerr{54.4}{+7.8}{-3.8}$ & $\asymerr{149}{+51}{-19}$ & ABDMG$^3$  & 31.9 $\pm$ 0.07  \\

J1633-6808B & M8.5 & $\asymerr{70.5}{+3.2}{-4.0}$ & $\asymerr{200}{+50}{-50}$ & CARN  &   30.3 $\pm$ 0.02  \\

J1839+4424B & M9 & $\asymerr{71.2}{+3.1}{-4.7}$ & $\asymerr{185}{+20}{-20}$ & Theia 213  &  30.8 $\pm$ 0.26  \\

J0458+1212 & L0.5 & $\asymerr{74.4}{+1.0}{-3.4}$ & $\asymerr{750}{+100}{-100}$ & HYA$^{6}$  &  18.8 $\pm$ 0.02  \\

J0410+1459 & L0.5 & $\asymerr{79.8}{+2.1}{-3.8}$ & $\asymerr{750}{+100}{-100}$ & HYA  &  29.4 $\pm$ 0.98  \\

J0442+1754 & L2-L3 & $\asymerr{80.1}{+1.6}{-3.5}$ & $\asymerr{750}{+100}{-100}$ & HYA  & 22.9 $\pm$ 0.70  \\

J1633-6808A & M8 & $\asymerr{84.4}{+9.0}{-12}$ & $\asymerr{200}{+50}{-50}$ & CARN   & 27.5 $\pm$ 0.06  \\

J1559+4404 & M8 & $\asymerr{88.8}{+4.8}{-4.6}$ & $\asymerr{162}{+2}{-2}$ & Oh 59$^{8}$  & 13.7 $\pm$ 0.04  \\

J1222+21401 & M7V & $\asymerr{92.9}{+1.7}{-4.1}$ & $\asymerr{414}{+23}{-23}$ & UMA$^{7}$  & 10.0 $\pm$ 0.25  \\

J1231+2416 & M7.3 & $\asymerr{122}{+4.0}{-7.6}$ & $\asymerr{562}{+98}{-84}$& CBER$^{9}$  & 6.27 $\pm$ 0.13  \\

J0426+1703 & M7V & $\asymerr{136}{+6.6}{-7.5}$ & $\asymerr{750}{+100}{-100}$ & HYA &  23.6 $\pm$ 0.04  \\

J1004+5022 & L3$\beta$ & - & - & Crius 227  & 25.3 $\pm$ 0.12  \\

J0219-3925B & L4$\gamma$ & - & $\asymerr{45}{+4}{-4}$ & THA  & 20.9 $\pm$ 0.04  \\

J0903-0637 & M7 & - & $\asymerr{45}{+15}{-5}$ & ARG  &17.5 $\pm$ 0.23  \\ 

J0609-3549 & L4 & - & $\asymerr{149}{+51}{-19}$ & ABDMG  & 6.27 $\pm$ 0.27  \\

J1131-3436A & M8.5/9 & - & $\asymerr{62}{+7}{-7}$ & CT$^{10}$ &  20.7 $\pm$ 0.23  \\

J2045-6332 & L1 & - & $\asymerr{162}{+2}{-2}$ & Oh 59   &  24.6 $\pm$ 0.06  \\

J2331-0406B & L3 & - & $\asymerr{149}{+51}{-19}$ & ABDMG  & -   \\ \bottomrule

\end{tabular}

\footnotesize{ \centering Age References : 
1 - \cite{2015MNRAS.447.1267M},  
2 - \cite{2018AAS...23132603Z}, 
3 - \cite{2015MNRAS.454..593B},
4 - \cite{2022ApJ...939...94M}
5 - \cite{2006ApJ...649L.115Z},
6 - \cite{2015ApJ...807...24B}, 
7 - \cite{2018ApJ...862..138G},
8 - \cite{2017AJ....153..257O},
9 - \cite{2014A&A...566A.132S},
10 - \cite{2020ApJ...903...96G} \\
\smallskip
Note: \textbf{OCTN} - Octans-Near association, \textbf{ARG} - Argus, \textbf{CAR} - Carina, \textbf{THA} - Tucana-Horologium association, \textbf{COL} - Columba, \textbf{TWA} - TW Hya, \textbf{CARN} - Carina Near, \textbf{ABDMG} - AB Doradus, \textbf{HYA} - Hyades Cluster, \textbf{UMA} - Ursa Major Corona, \textbf{CBER} - Coma Berenices, \textbf{CT} - Cas-Tau Association}

\end{table}

\section*{ACKNOWLEDGEMENTS} This study was supported by the Science and Technology Facilities Council through a PhD studentship to SB (ST/W507490/1) and through research infrastructure support to SB and HRAJ (ST/V000624/1). RLS has been supported by a STSM grant from  COST Action CA18104: MW-Gaia. We also acknowledge the valuable contributions of TOPCAT \citep{2005ASPC..347...29T, 2006ASPC..351..666T}, Scipy \citep{2020SciPy-NMeth}, Astropy \citep{2013A&A...558A..33A}, the VizieR catalogue access tool and the SIMBAD database operated at CDS, Strasbourg, France; National Aeronautics and Space Administration (NASA) Astrophysics Data System (ADS). The $\gaia$ mission has been pivotal for this work provided by the European Space Agency (ESA) \href{https://www.cosmos.esa.int/gaia}{(https://www.cosmos.esa.int/gaia)} processed by the Gaia Data Processing and Analysis Consortium (DPAC, \href{https://www.cosmos.esa.int/web/gaia/dpac/consortium}{(https://www.cosmos.esa.int/web/gaia/dpac/consortium)}. Funding for the DPAC has
been provided by national institutions, in particular, the institutions participating
in the Gaia Multilateral Agreement. This research made use of the cross-match service provided by CDS, Strasbourg. We thank the anonymous referee for their thorough examination and suggestions.
We also extend our gratitude to Dr Ben Burningham and Dr Alessandro Sozzetti for their valuable feedback.

\section*{DATA AVAILABILITY}
The UCDC will be hosted at CDS. It can also be accessed at \href{https://zenodo.org/records/13312178}{https://zenodo.org/records/13312178}.

\bibliographystyle{mnras}
\bibliography{Binary.bib}

\bsp	
\label{lastpage}
\end{document}